\documentclass[useAMS,usenatbib]{mn2e}
\usepackage{graphicx}
\usepackage{mathrsfs}
\usepackage{amssymb}
\usepackage{amsmath}
\usepackage{mathtools}
\usepackage{xcolor}
\usepackage{txfonts}
\usepackage{ulem}
\usepackage{framed}
\usepackage{lastpage}
\usepackage{tabularx}
\DeclareSymbolFont{cmletters}{OML}{cmm}{m}{it}
\DeclareMathSymbol{v}{\mathalpha}{cmletters}{"76}
\title[Variability in black hole accretion flows]{Effects of adiabatic index on the sonic surface and time variability of low angular momentum accretion flows}
\author[Palit, Janiuk, Sukova]{ Ishika Palit $^{1}$, Agnieszka Janiuk $^{1}$, Petra Sukova $^{2}$ \\
$^1$Center for Theoretical Physics, Polish Academy of Sciences, Al. Lotnikow 32/46, 02-668 Warsaw, Poland\\
$^2$ Astronomical Institute, Czech Academy of Sciences, Bo\v{c}n\'{i} II 1401, CZ-141 00 Prague, Czech Republic\\
}

\begin{document}

\date{}
\pagerange{\pageref{firstpage}--\pageref{lastpage}} 
\pubyear{}
\maketitle
\label{firstpage}

\begin{abstract}
We study the role of adiabatic index in determining the critical points in the transonic low angular momentum accretion flow onto a black hole. We present the general relativistic 2D hydrodynamic simulations of  axisymmetric, inviscid  accretion flows in a fixed Kerr black hole gravitational field. A relativistic fluid where its bulk velocity is comparable to the speed of light, flowing in the accretion disk very close to the horizon can be described by an adiabatic index of $4/3 < \gamma < 5/3$. The time dependent evolution of the shock position and respective effects on mass accretion rate and oscillation frequency with varying adiabatic index is discussed in the context of the observed microquasars.
\end{abstract}

\begin{keywords}
accretion, accretion discs, black hole physics, shock waves, hydrodynamics, MHD, stars:X-rays:binaries, QPOs
\end{keywords}

\maketitle

\section{Introduction}
When a massive star is in the late stages of evolution, it undergoes a gravitational collapse that results in an extremely violent supernova explosion and the formation of a black hole. A good amount of stars in the sky are in the so-called 'binary systems' where two stars orbit around one another. If the companion star survives the explosion, a star-black hole system is formed, called an X-ray binary.
In X-ray binaries, material coming from the companion star is spiralling onto the black hole, forms an accretion disc, where very hot matter (about 10 million K) heats up and produces X-ray radiation due to intense frictional forces. It is thought that, in addition, there is an even hotter plasma (about 10 billion K) that sandwiches the disc around its inner regions, which we call the 'corona' \citep{kawabata2010radiative}. 

The well known fact about black-hole X-ray binaries is that they  show two basic spectral states: a high/soft state appearing at high luminosities and a low/hard spectral state at low luminosities \citep{belloni2011black}.
In the inner hot corona, the accretion proceeds at lower rates, and is described by models such as the advection-dominated accretion flow  \citep{ichimaru1977bimodal, rees1982ion, chen1995unified}. 
Models of hot accretion flows have been studied in detail since very long time but its completeness is rendered due to some theoretical uncertainties on heating of electrons, equilibration of electron and ion temperatures, and relative roles of thermal and non-thermal particles \citep{esin1996hot, ozel2000hybrid, veledina2013hot}. Currently various kinds of models are being explored via hydrodynamic and magneto-hydrodynamic computer simulations in order to study the link between hot accretion flow and their outflows as theory suggests strong wind production for these kind of accretion flows \citep{hawley2001magnetohydrodynamic, igumenshchev2003three, narayan2000self, waters2018magnetothermal}.

Due to advection of energy, the radiative efficiency of these flows is lower than that of a standard thin accretion disk. The observed spectrum of a black hole accretion disk has been anticipated to be coming in the form of multi-colour black body emission \citep{shakura1973black, abramowicz1988slim} and the power-law component from a centrifugal force dominated boundary layer (CENBOL). Detailed description of such two Component Advective Flow (TCAF) model is known for a long time, as mentioned in \citep{chakrabarti1995spectral, chakrabarti1997spectral}; also see \cite{cabanac2010variability}  where they show the variability of X-ray binaries from an oscillating hot corona.

In TCAF model, a Keplerian disk with high viscosity at the equatorial plane is present inside a low angular momentum sub-Keplerian halo \citep{kumar2014dissipative}.  
This part emits flux of radiation similar to the one calculated for  Shakura - Sunyaev  disk. 
In the two-component advective flow (TCAF) model, the soft photons coming from the Keplerian disk get intercepted by  post-shock region and leave the system as hard radiation. The oscillations of the post-shock region will leave their signature in the outgoing hard photons \citep{chakrabarti1995spectral}.

Inviscid flow of low angular momentum matter falling onto the black hole, feels the centrifugal barrier (scaling as $ ~ {l^ {2} /r^{3}}$, l and r being the specific angular momentum and the radial distance). This would slow it down, and pile up eventually making a possible density jump, before entering the black hole. In viscous flows, the result depends on the exact magnitude of viscosity. These shock waves just outside the horizon provide an opportunity to study the emission of radiation in a strong gravity limit and would therefore be of great interest, especially to pinpoint the mass and the spin of the central object. In presence of high accretion rates in the Keplerian component, the CENBOL can be cooled down so that the spectrum may be totally dominated by the low energy X-rays \citep{popham2001accretion, white1988x}. The spectrum goes to a so-called 'soft state'. When the Keplerian rate is not high compared to the low-angular momentum component, the CENBOL survives and the spectrum is dominated by the high energy X-rays. Then it is said to be in a so-called 'hard-state'. In the presence of radiative or thermal cooling effects, CENBOL may start to oscillate, especially when the infall time scale and the cooling time scale are comparable. In this case, number of intercepted low energy photons would be modulated. As a result, the number of high energy photons are also modulated. This effect produces what is known as quasi-periodic oscillations (or QPOs) in black hole candidates \citep{chakrabarti1997spectral}.
\\
It is not viable to describe black hole accretion by a fixed $\gamma$ equation of state (EoS) from infinity to horizon. For non-relativistic regime where $\frac{kT}{mc^{2}} << 1$ , EoS assumed with the adiabatic index of $\gamma $ = 5/3 can be appropriate.  For ultra-relativistic where $\frac{kT}{mc^{2}} \geq 1$ the flow needs to be  described by relativistic Eos with $\gamma$ = 4/3, rather than ideal gas EoS. Here k, T and m represents the Boltzmann constant, temperature and mass of the constituent of the flow correspondingly. For example, weakly active galaxies, such as Sgr A, usually studied using a polytropic Eos with $\gamma$ around 5/3, because these flows are believed to be radiatively inefficient and gas pressure dominated. 
The other scenario could be imagined of GRB's which are a blend of electron-positron pairs, nucleons, photons and neutrinos \citep{janiuk2004evolution} where the accretion flow is radiation pressure dominated \citep{meszaros2006gamma}, and requires a relativistic EoS with $\gamma$ = 4/3.
Also, even smaller values of adiabatic index, close to isothermal value of 1.0,  may be relevant to the systems like the proto-planetary and proto-galactic disks \citep{janiuk2009time}. \\
Many authors have investigated  the inviscid accretion models in general relativistic regime with relativistic equation of state \citep{fukue1987transonic, chattopadhyay2009effects}. The general outcome of all the past analysis shows that the transonic, radiative and thermodynamic behaviour
of the flow is strongly related to its composition.
Adiabatic index gives away the information about microphysics of the flow and thus different values of $\gamma$ correspond to different types of astrophysical objects or different phases of astrophysical activity. Among the three parameters determining the sonic points and shock position in the transonic solution, adiabatic index also seems to play a crucial role in determining the formation of standing or oscillating shocks.\\
In former work, \cite{sukova2017shocks}, the numerical studies of quasi-spherical transonic accretion and time-dependent evolution of the shock position were for the first time addressed in a full General Relativistic scheme. This is why the results obtained were not only qualitatively, but also quantitatively relevant for the observation of realistic sources, where the GR effects cannot be neglected in the closest vicinity of the black holes. That work however was limited to the choice of a fixed parameter, namely the adiabatic index of $\gamma=4/3$, as is relevant for the radiation pressure dominated ideal equation of state. In the current work, we expand the parameter space of these simulations, and we study the role of adiabatic index in producing different pattern of shock evolution.\\
This paper is organized as follows. In Section \ref{sec:model} we present our initial conditions and explain the evolution of the flow over time. The subsections contain the description of our computational and numerical scheme. In Section \ref{sec:results} we present the results from our simulation as how the shock behaves for different models, its effect with changing adiabatic index and for spinning black hole.
In Section \ref{sec:discussion}, we discuss our results and present analogy with observations of microquasars and give conclusions.


\section{Model}
\label{sec:model}

\subsection{Initial Conditions}
To study the accretion flow of non-viscous matter, we start with the polytropic equation of state:
\begin{align}
p = K \rho ^{\gamma}
\end{align}
where $\gamma $ is the adiabatic index, p is the gas pressure and $\rho $ is the gas density. 
The local sound speed, $u_{s}$, is given by the relation
\begin{align}
    u_{s}^{2} = \frac{\gamma p}{\rho} = \gamma K \rho^{\gamma-1}.
\end{align}
Unlike for a thin disc, here  we assume the quasi-spherical distribution of the gas, provided by constant specific angular momentum $\lambda$ \citep{abramowicz1981rotation}. Such a distribution of matter is possible to be formed instead of an evaporated Keplerian accretion disc.\\
A transonic flow of matter makes a jump from supersonic to subsonic region at certain position, which is called shock ($r_{s}$) \citep{chakrabarti1989standing}.  The region from shock to inner sonic point is being defined as the post shock region, which basically acts as CENBOL region.
The Rankine-Hugoniot conditions have to be fulfilled at the shock position $r_{s}$. Using those conditions, expression for Mach number ($ \mathscr{M}$) is
\begin{equation}
\frac{(\frac{1}{\mathscr{M}_{+}} + \gamma \mathscr{M}_{+})^{2}}{\mathscr{M}_{+}^{2}(\gamma - 1)+2}    = \frac{(\frac{1}{\mathscr{M}_{-}} + \gamma \mathscr{M}_{-})^{2}}{\mathscr{M}_{-}^{2}(\gamma - 1)+2} 
\end{equation}
 where $\mathscr{M} = \frac{(v)}{(u_{s})}$  i.e dimensionless ratio of flow velocity at a boundary to the local sound speed.\\
The increase in density across the shock is described by compression ratio, $\mathcal{R}_{comp} = \rho_{+}/\rho_{-}$ , where $\rho_{+}$ is post shock density and $\rho_{-}$ is pre shock density. 

The initial conditions that we use here to study the $\gamma$ dependence of the sonic surface are similar to those used in previous work \citep{sukova2015shocks}.
The rotation, i.e. the angular momentum of the flow has been prescribed according to the relation 
\begin{align}
  \lambda = \lambda^{eq} \sin^{2}\theta, 
  \label{eq:lambda}
\end{align} 
where $\theta = \pi /2 $ and $\lambda^{eq}$ is the angular momentum in the equatorial plane.  The definition of angular momentum goes as usual,  $\lambda = \frac{u^{\phi}}{u^{t}}$.\\

Our initial state does not correspond to the stationary state, because that is derived for quasi-spherical distribution of gas with constant angular momentum in the pseudo-newtonian potential. However, as we scale angular momentum according to relation (\ref{eq:lambda}) and we are in the general relativistic regime, the resulting configuration is not the solution of the stationary time-independent equations. Hence, it is expected that at the beginning of the simulation during a transient time, the flow adjust itself into the appropriate profile.

In our computations, the variables adopted initially in the function of radius, are $\rho$, $\epsilon$  with $\epsilon = K \rho^{\gamma-1} / \gamma -1$.  Here K denotes entropy and is given by 
\begin{equation}
\textrm K = \left \{ vr^{2}\frac{u_{s}^{\frac{2}{\gamma-1}}}{\gamma^{\frac{1}{\gamma-1}} \dot{\textrm M} }\right \}^{\gamma - 1}.
\end{equation}
The radial gradient for flow velocity must be real and always finite to maintain the smooth and continuous accretion flow from outer sonic point to inner sonic point.
Critical point of the flow has been found by setting numerator and denominator of the velocity gradient = 0. Thus equation for critical point position comes as:
\begin{equation}
\epsilon - \frac{\lambda^{2}}{2r_{c}^{2}} - \frac{\gamma+1}{2(\gamma-1)}(\frac{r_{c}}{4(r_{c}-1)^{2}} - \frac{\lambda^{2}}{2r_{c}^{2}} ) = 0 
\end{equation} 
and the velocity gradient at critical point is obtained as :
\begin{align}
      \frac{dv}{dr}\mid_{r_{c}} =  \frac{-B \pm C}{2A}
      \label{eq:gradient}
\end{align}
where \\
B = $4\frac{a^{2}(\gamma-1)}{v\cdot r_{c}}$, \\ [1ex]
C = $\sqrt{(-4\frac{a^{2}(\gamma-1)}{v\cdot r_{c}})^{2} -4(1+\frac{\gamma a^{2}}{v^{2}})(\frac{3\lambda^{2}}{r^{4}} -\frac{1}{(r_{c}-1)^{3}} +\frac{2a^{2}(2\gamma-1)}{r_{c}^{2}}))}$, \\[1ex]
A = $(1+\frac{\gamma a^{2}}{v^{2}})$. \\ 

The models are parameterized by the value of the specific angular momentum $\lambda $, polytropic exponent $\gamma$ and the energy $\epsilon$, which set the critical point position $r_{c}^{in}$, $r_{s}$ $r_{c}^{out}$  \citep{chakrabarti1996accretion}.

 We get two values for velocity gradient at critical point in equation \ref{eq:gradient}, which are obtained as real and opposite to each other. This shows that the nature of the critical points are saddle type.


\subsection{Time evolution of the flow}
\label{sec : time evolution}

The quasi spherical, slightly rotating flow in all the models in our simulations, starts with the initial condition prescribing the critical point  and the velocity gradient at the critical point derived in \cite{sukova2015shocks}. The model parameters i.e the specific energy ($\epsilon$), specific angular momentum ($\lambda$), adiabatic index ($\gamma$), spin of black hole (a),  inner and outer radius of the computational grid  R$_{out}$ and R$_{in}$, resolution and distribution of angular momentum are chosen in the initial condition, which set the properties of the flow. The resolution for all the models has been chosen as [384*256] in radial and theta direction.\\
The flow evolves and shock position can be seen oscillating, expanding or accreting in response to pressure against rotational force.
The end time of the 2D simulations for all the models presented here is $t = 10^{6}$ [M] (here t [M] = GM$_{\textrm BH}$/c$^{3}$ converts the code units to physical timescale, so as a result we have t$_{\rm final}= 49.27$ sec for a 10 M$_{\odot}$ black hole). The units for in the  numerical simulation for the models are in geometrical units $(G = c = 1, [r] = [t] = [\lambda] =$[M]).
The mass accretion rate has been calculated as in \cite{gammie2004black}:
\begin{equation}
    \dot{\rm M}(r,t) = \iint \rho u^{r} \sqrt{-g} d\theta d\phi
\end{equation}
The value of accretion rate presented here are in code units. They can be converted into physical units for physical realization. The units of density and time have to be adopted  for conversion where the latter quantity follows from the assumed black hole mass value, and hence the light crossing time. The critical mass accretion rate is
\begin{equation}
     \dot{\rm M}_{\rm Edd} = \frac{\rm L_{\rm Edd}}{\eta c^{2}} = 1.39 \times 10^{18} \frac{M}{M_{\odot}} ~~ [\rm g s^{-1}]
\end{equation}
where  L$_{Edd}$ is Eddington limit of luminosity, and $\eta$ is around 0.1 for transonic accretion \citep{fukue2004critical}.
Our chosen models of simulation should satisfy $\dot{\rm M} < \dot{\rm M}_{\rm Edd}$
(see Table 1 from \cite{janiuk2011different} for expected mass accretion rate for low mass X ray binaries and microquasars in terms of $\dot{\rm M}_{Edd}$ ).


\subsection{Details of numerical scheme}

The evolution of non-magnetized gas (as assumed in our initial conditions) is simulated with the HARM package supplied with a few modifications (see Sukova et al. 2017 for details).
The code conserves the vanishing magnetic field and there is no spurious magnetic field generated during the evolution.

HARM (high-accuracy relativistic magnetohydrodynamics), is a conservative, shock-capturing scheme for evolving the equations of general relativistic MHD (\citep{gammie2003harm}).
The fundamental equations used in HARM are:
particle number conservation equation
\begin{equation}
    (nu_{\mu});_{\mu} = 0,
\end{equation}
the four energy momentum equations (in coordinate basis):
\begin{equation}
    \frac{1}{\sqrt{-g}} \delta _{\mu}(\sqrt{-g}\rho u^{\mu}) = 0
\end{equation}
MHD stress energy tensor conservation: 
\begin{equation}
    T^{\mu}_{\nu};\mu = 0
\end{equation}
and, in case of magnetized flow, the induction equation (in coordinate basis)
\begin{equation}
    \delta_{\mu}(-gB^{i}) = -\delta_{j}(-g(b^{j}u^{i} - (b^{i}u^{j} )).
\end{equation}

These hyperbolic, GRMHD equations are written in a conservative form in HARM code to integrate them numerically.
Being a conservative scheme, its needed to update a ‘‘conserved’’ variable at each time step using fluxes. The vector for conserved variables used in HARM is 
\begin{equation}
    U = \sqrt{-g} (\rho \textit{u}^{t}, T^{t}_{t}, T^{t}_{i}, B^{i}).
\end{equation}
To model the flow, code needs to make a choice of primitive variables to be interpolated within zones.
Such primitive variables in the code have simple physical interpretation such as density, velocity and magnetic field (B = 0 for our non magnetized fluid)
\begin{equation}
    P = (\rho, \textit{u}, \textit{v}^{i}, B^{i}).
\end{equation}
As U is updated rather than each primitive variable, code calculates U(P) at the end of each timestep using the value of P from the last timestep as an initial guess for a multidimensional Newton-Raphson routine. The calculation of the Jacobian ($\delta U/\delta P$) involved here is computationally expensive task.
\\ [1.5ex]
The initial conditions are set using Boyer-Lindquist coordinates, and they are transformed into the code coordinates, which are the Kerr-Schild ones (see \citep{weinberg1973gravitation, boyer1967maximal, visser2007kerr}).
The code HARM used here in this paper to study the accretion disk has been modified to
produce the Jacobian transformation from code coordinates to any desired coordinates as of the metric used. 
The transformation matrix has been coded into the coordinate file of the scheme and has been dumped in the output file which helps to use the primitive variables such as the four-velocities, in the Kerr-Schild (KS), modified Kerr-Schild (MKS) as well as Boyer-Linquidst (BL) coordinates. 
The KS - MKS coordinates are provided by the transformation from Kerr Schild’s coordinates (t, r, $\theta$, $\phi$) to the modified KS (t$^{[MKS]}$, x$^{[1]}$, x$^{[2]}$, x$^{[3]}$)  was proposed in \cite{gammie2003harm}. The transformation of the coordinate follows as:
\begin{equation}
  t =  t^{[MKS]}
\end{equation}
\begin{equation}
    r = e^{x^{[1]}} + constant
\end{equation}
\begin{equation}
    \theta = \pi x^{[2]} + \frac{1 - h}{2} \sin(2\pi x^{[2]})
\end{equation}
\begin{equation}
    \phi =  x^{[3]}
\end{equation}
We use $h = 1.0$ here for uniform angular grid and the $\theta $ coordinate is simply re-scaled by $\pi$. 



\subsubsection{Computational setup and grid}
 We have computed  set of models, spanning the parameter space $(\gamma,\epsilon,\lambda)$ for shock formation and sonic behaviour of accretion flow. They are presented in Table [\ref{Table 1}] with the initial condition given by the shock solution for non-rotating and rotating black holes.
 Six values of $\gamma $ have been considered  describing different gas micro-physics: $:1.02,1.2,4/3,1.4,1.5,5/3$. The runs were performed  for two  values of energy and four values of angular momentum, which were lower and higher than the critical values defined in \cite{sukova2017shocks}.

 To achieve the fine resolution for the whole accretion structure, the grid has been set to logarithmic in radius near the black hole and as super exponential grid spacing in outer region. 
 In the innermost region, the grid spans below the horizon as MKS coordinate log(r) is used as radial coordinate instead of usual Boyer-Lindquist r. This concentrates numerical resolution toward the
horizon and the regularity of KS coordinate makes it possible to have several zones inside the black hole and the free outflow boundary. 

 It seems also important to have supply of matter throughout the time of evolution of flow (otherwise all the gas gets accreted leaving empty density profile). The inflow of matter (its density and velocity) through the outer boundary is given by the stationary solution for the appropriate radius of the two outer ghost zones at the initial time, so it mimics the prolongation of the initial density and velocity outwards, and it stays constant during the evolution. This enables us to simulate the flow behaviour on long time scale.

\section{Results}
\label{sec:results}
The results for all the models are summarized in the Table [\ref{Table 1}]. The table contains information about model parameters and if the shock was found, we give the information about its nature (accreting, expanding or oscillating). We also give the value of the shock radius and compression ratio. The models B1[$\gamma$ = 1.2 , $\lambda$ = 3.6 [M] , $\epsilon$ = 0.0025], B2[$\gamma$ = 1.2 , $\lambda$ = 3.8 [M] , $\epsilon$ = 0.0025], B3[$\gamma$ = 1.2 , $\lambda$ = 3.58 [M] , $\epsilon$ = 0.0005], D1[$\gamma$ = 1.4 , $\lambda$ = 3.6 [M] , $\epsilon$ = 0.0025], D2[$\gamma$ = 1.4 , $\lambda$ = 3.8 [M] , $\epsilon$ = 0.0025], E1[$\gamma$ = 1.5 , $\lambda$ = 3.6 [M] , $\epsilon$ = 0.0025], E2[$\gamma$ = 1.5 , $\lambda$ = 3.8 [M] , $\epsilon$ = 0.0025] and E3[$\gamma$ = 1.5 , $\lambda$ = 3.58 [M] , $\epsilon$ = 0.0005] are not listed in the Table because there was no shock formation and there existed only one sonic point for these models.
Models C1[$\gamma$ = 4/3 , $\lambda$ = 3.6 [M] , $\epsilon$ = 0.0025], C2[$\gamma$ = 4/3 , $\lambda$ = 3.8 [M] , $\epsilon$ = 0.0025], C3[$\gamma$ = 4/3 , $\lambda$ = 3.58 [M] , $\epsilon$ = 0.0005] and C4[$\gamma$ = 4/3 , $\lambda$ = 3.72 [M] , $\epsilon$ = 0.0005] are also not presented in the Table [\ref{Table 1}]. These models can be found in \cite{sukova2017shocks}.

\begin{figure}
      \includegraphics[width = 0.5\textwidth]{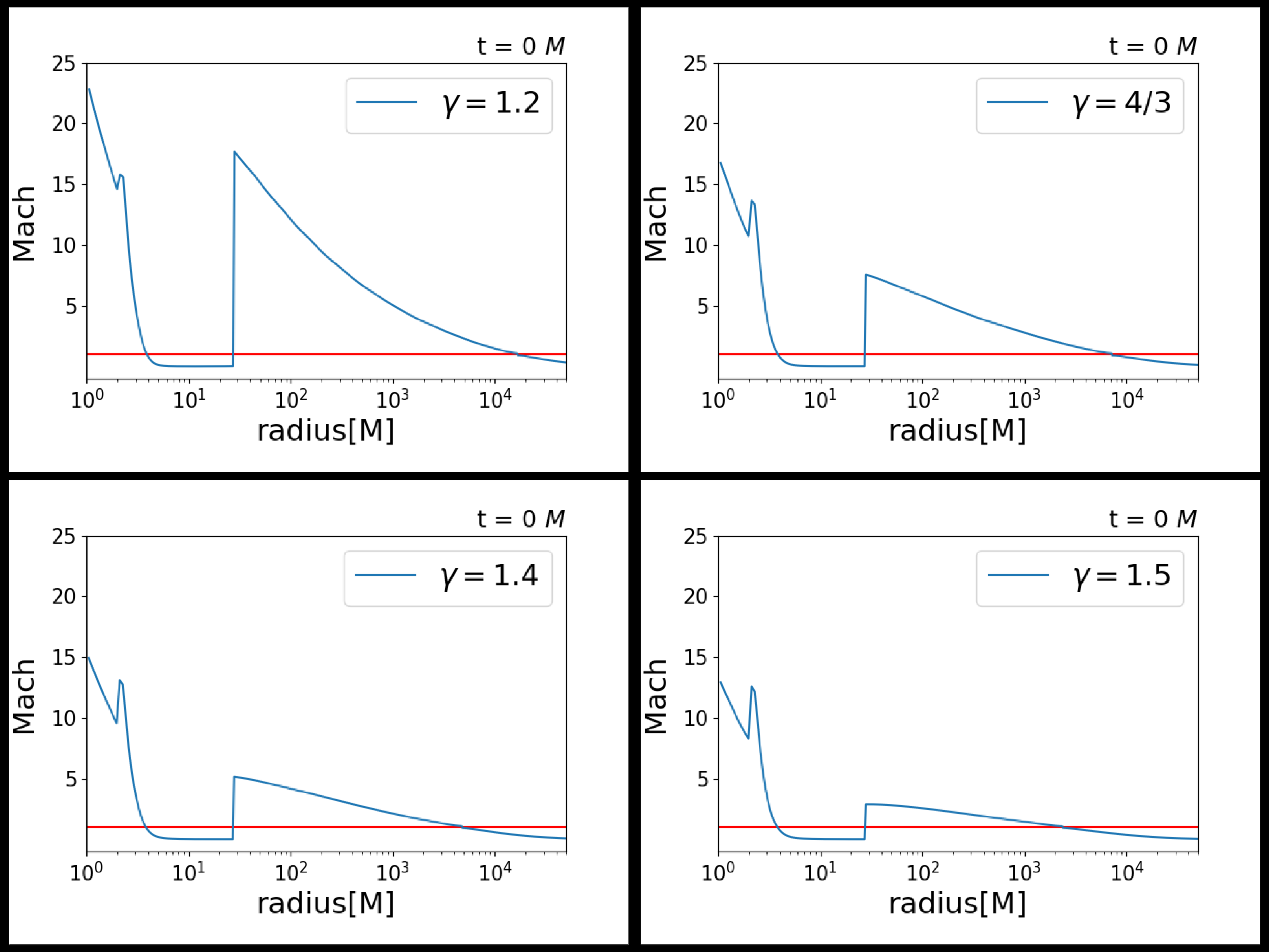}
     \caption{The 1D Mach number calculated over the equator has been shown here with all the critical points for models with different adiabatic index but the same  $\epsilon$ = 0.001 and $\lambda$ = 3.86 [M]. In the simulation, $\gamma = 1.2$ corresponds to models named with 'B',$\gamma = 4/3$ corresponds to models named with 'C',  $\gamma = 1.4$ corresponds to models named with 'D',  $\gamma = 1.5$ corresponds to models named with 'E'. See text for details. }
     \label{fig:1}
 \end{figure}

Figure [\ref{fig:1}] shows the  Mach number corresponding to different values of adiabatic index at the very beginning of the evolution for models B5[$\gamma$= 1.2], C5[$\gamma$= 4/3], D5[$\gamma$= 1.4] and E5[$\gamma$= 1.5] with same $\epsilon$ = 0.001 and $\lambda$ = 3.86 [M]. It can be inferred from here that as the flow velocity decreases with increasing value of $\gamma$ results in lower Mach number. Also the outer sonic point is located very far for lower adiabatic index.

Below in Sections \ref{sec:shock_behaviour} and \ref{sec:dependence}, we present the results of simulations and we illustrate them with time snapshots of the structure of the flow for chosen models. Every snapshot in Figures \ref{fig:2}, \ref{fig:4}, \ref{fig:6}, \ref{fig:7}, \ref{fig:8} and \ref{fig:9}, contains four panels with the slices of Mach number, $\lambda$ in geometrized units, and $\rho$ in arbitrary units. Figures are labelled by the time [t] in geometrized units, 
The axes show the position in geometrized units. 
We also show the distribution of Mach number in the equator.
In the Mach profile, the red colour corresponds to the supersonic motion and blue regions indicate the subsonic accretion. 

\subsection{Shock behaviour}
\label{sec:shock_behaviour}

\begin{figure}
       \textbf{ t = 0[M]}\\
     \includegraphics[width = 0.5\textwidth]{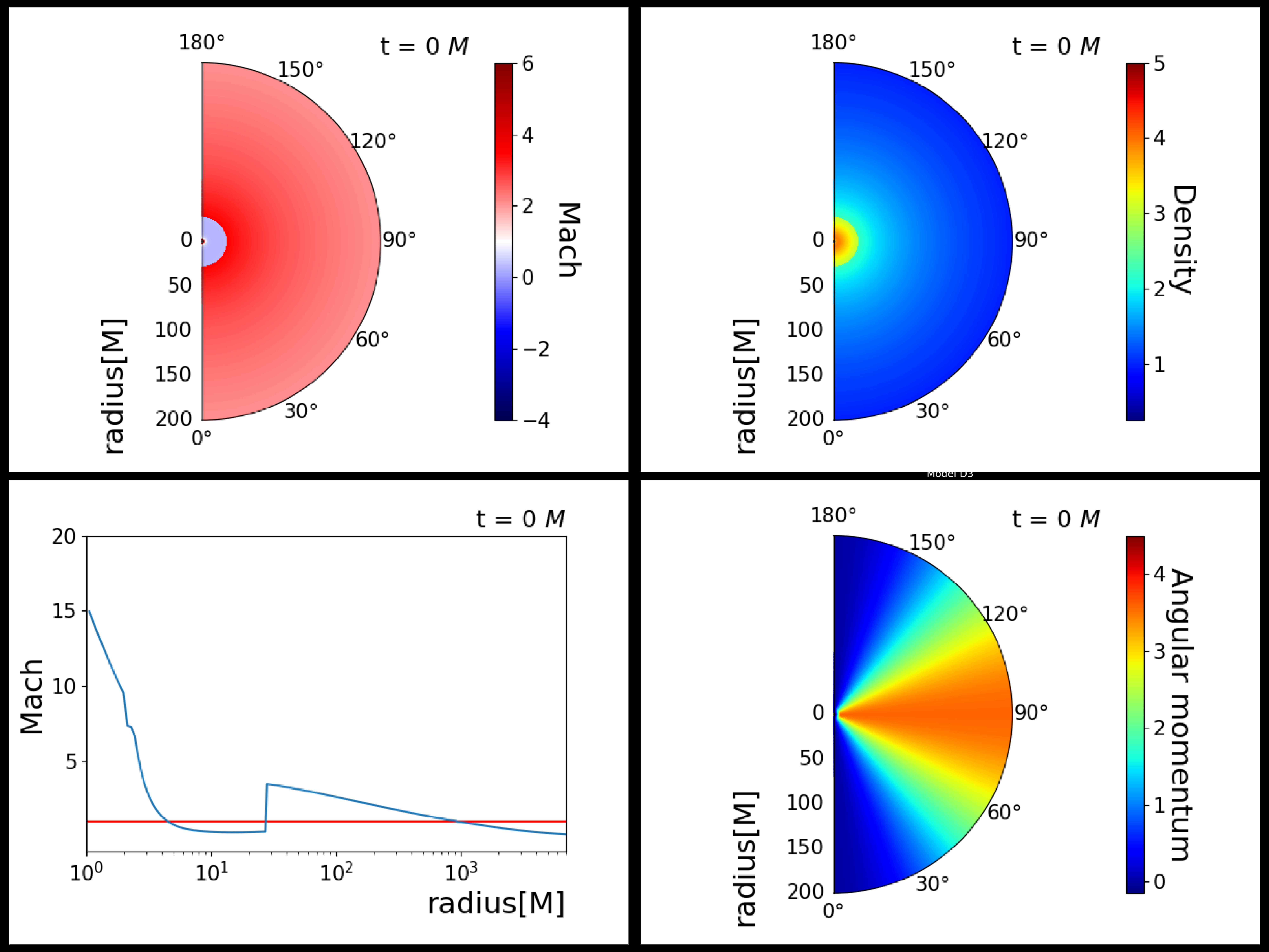}
     \caption{Models D3$[\gamma = 1.4, \lambda =3.58 \rm [M], \epsilon = 0.0005 ]$ at the time  $t = 0$ [M] showing the initial Bondi like condition for Mach number, angular momentum and density profile. The 1D plot of Mach number along the equatorial plane shows the position of outer sonic point, shock and inner sonic point. }
     \label{fig:2}
 \end{figure}

Figure [\ref{fig:2}] shows the initial condition for the model D3.
We obtain the solution of the flow and the values of transonic points namely inner and outer sonic point which are consistent with the analytical solution. These points are located at r$_{c}^{in}=$ 4.37 [M], r$_{c}^{out}=$917.7[M], and r$_{s}=$27.3[M].\\
The simulation of models where shock can appear shows that it can either :\\(i) stay at certain position,\\ (ii) oscillate in time,\\(iii) be accreted quickly from the the minimal stable shock position (close to the black hole),\\ or \\(iv) be formed close to the black hole and expand quickly through the outer sonic point.\\

The shock converts supersonic gas into denser, slower moving, higher pressure, subsonic gas hence it
increases the specific entropy of the gas. Above inner sonic point, where gas is dominantly subsonic, the shock dissociates molecules, or raises the temperature so that previously inaccessible degrees of freedom become accessible. This leads to different shock and sonic positions for different $\gamma$.\\

\begin{table*}
\centering
\begin{tabular}{c|c|c|c|c|c|c|c|c|c}
   \large{Models}& \large{$\gamma$}& \large{$\lambda^{eq}[M]$}& \large{$\varepsilon$}& \large{BH spin (a)}& \large{shock}& \large{$r_{s} [M]$}& \large{Mean $r_{s}$}&  \large{Compression ratio ($\mathcal{R}$})& \large{figures} \\ [2ex] 
  \hline 
    \hline A1& 1.02& 3.86& 0.0001&  0& No& -& -& -& - \\ [1ex]
   \hline  B4& 1.2& 3.72& 0.0005&  0& No& -& -& -& - \\ [1ex]
   \hline  B5& 1.2& 3.86& 0.0001& 0& OS$_{vs}$ + AC& 12.8 -- 90.4& 51.6& 1.18& (\ref{fig:4}, \ref{fig:5})  \\ [1ex]
    \hline  B6& 1.2& 3.6& 0.0001&  0& AC& -& -& -& - \\ [1ex]
   \hline  C5& 4/3& 3.86& 0.0001&   0& OS&  25.4 -- 314& 169.7& 3.6& (\ref{fig:1}) \\ [1ex]
   \hline C6& 4/3& 3.6& 0.0001&  0& AC& -& -& -& - \\ [1ex]
    \hline D3& 1.4& 3.58&  0.0005&  0& OS& 99.9 -- 151& 125.5& 2.9& (\ref{fig:2},\ref{fig:6})\\ [1ex] 
    \hline D4& 1.4& 3.72& 0.0005&  0& OS& 101 --1033& 567&  1.8 & -\\ [1ex]
   \hline  D5& 1.4& 3.86& 0.0001&  0& OS$_{vs}$& 10.2 -- 3789& 1900& 1.7& (\ref{fig:7}, \ref{fig:14})\\ [1ex]
   \hline  D6& 1.4& 3.6& 0.0001&  0& OS& 44.3 -- 97.5& 70.9&  3.01& (\ref{fig:8},\ref{fig:12})\\ [1ex]
   \hline  E4& 1.5& 3.72& 0.0005&  0& AC& -& -& -& - \\ [1ex]
  \hline   E5& 1.5& 3.86& 0.0001& 0& EX& 34.9& -& 2.7& (\ref{fig:9}, \ref{fig:15}b)\\ [1ex]
  \hline   E6& 1.5& 3.6& 0.0001&  0& EX& 92.7& -&  2.9& (\ref{fig:10})\\ [1ex]
  \hline   F1& 5/3& 3.58& 0.0005& 0& No& -& -& -& -  \\ [1ex]
  \hline  H1& 1.4& 3.86& 0.0001&  0.10& OS$_{vs}$ + Ex& 68.3& -& 2.04& -  \\ [1ex]
    \hline   H2& 1.4& 3.86& 0.0025& 0.10& AC&  -& -& -& -  \\ [1ex]
     \hline   H3& 1.4& 3.86& 0.0001& 0.89& Ex& 2390& -& 2.95& -  \\ [1ex]
      \hline   H4& 1.4& 3.6& 0.0001& 0.10& OS&  58.7 - 219.6& 139.2& 2.34& (\ref{fig:11},\ref{fig:13})  \\ [1ex]
  \hline   H5& 1.4& 3.6& 0.0025& 0.10& AC& -& -& -& -  \\ [1ex]
  \hline   H6& 1.4& 3.6& 0.0001& 0.89& EX&  1106& -& 2.15& -  \\ [1ex]

\end{tabular}\caption{The columns from $(1)-(10)$ show: the name of the run model, adiabatic index, specific energy, specific angular momentum, the black hole spin for each model, the nature of shock surface, position of shock appearance or oscillation, mean position of shock oscillation, compression ratio and  figures related to the models presented in the article here. Here AC denotes accreting nature of the shock through inner sonic point, EX denotes expanding nature of the shock through outer sonic point, and OS denotes oscillating shock. OS$_{vs}$ denotes very small oscillation of the shock position for a short period of time. Compression ratio is calculated at the shock position close to the horizon. The resolution for all the models  is $384*256$ (see Section \ref{sec : time evolution}). }
\label{Table 1}
\end{table*}

 \begin{figure}
    \centering
     \textbf{(a)}\\
     \includegraphics[width = 0.5\textwidth]{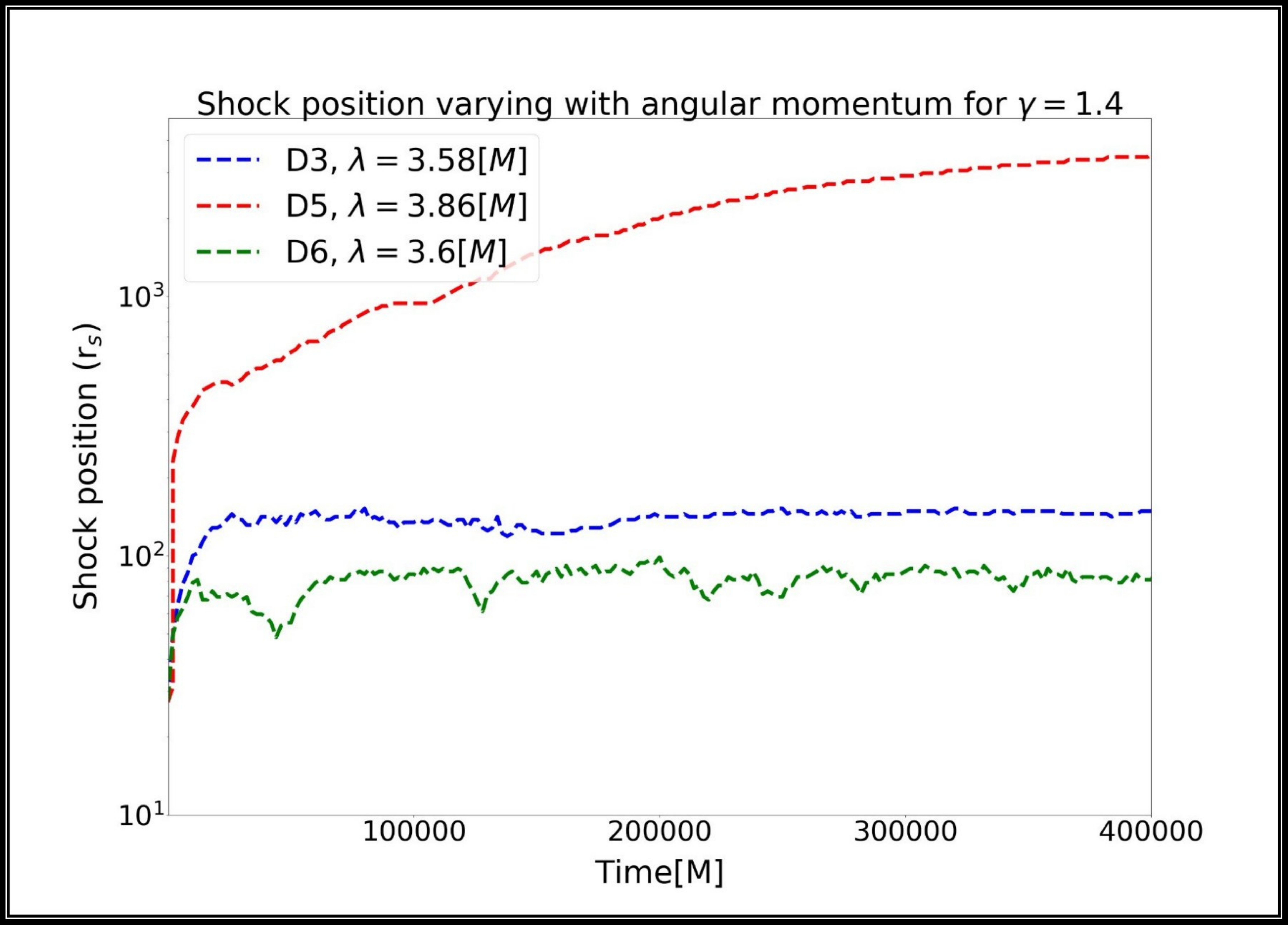}\\
     \textbf{(b)}\\
    \includegraphics[width = 0.5\textwidth]{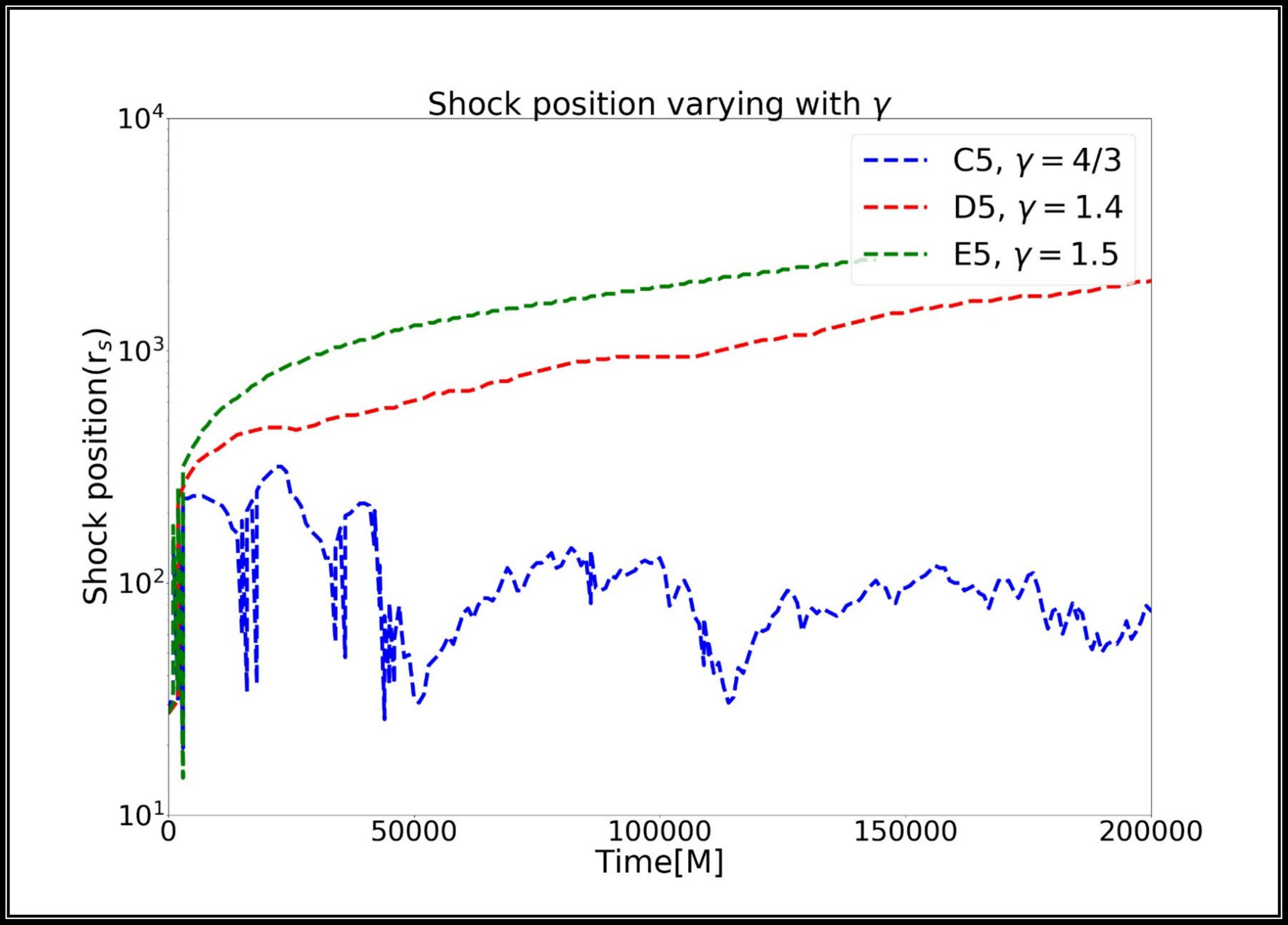}\\
     \textbf{(c)}\\
     \includegraphics[width = 0.5\textwidth]{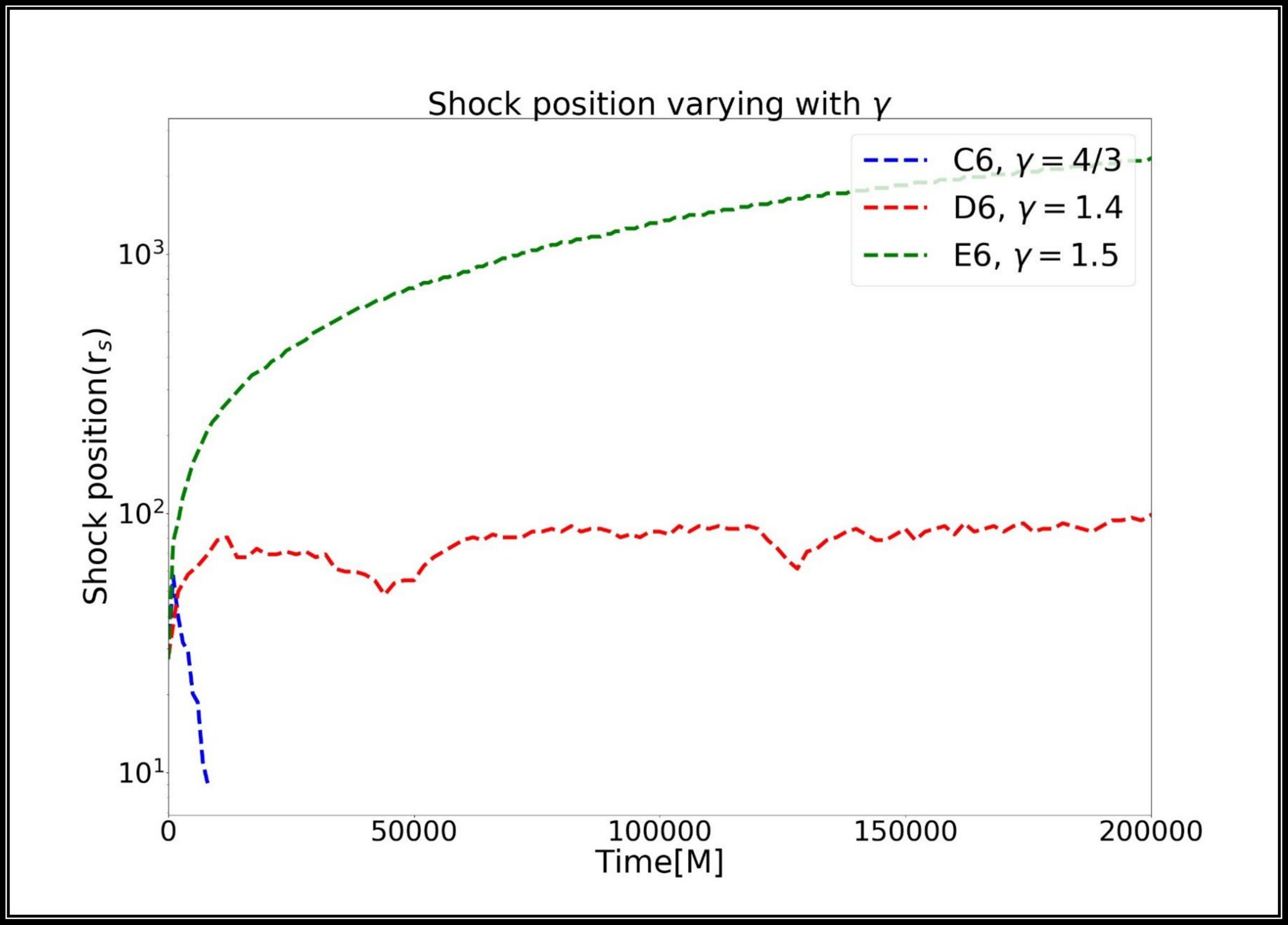}\\
     
     \caption{Panel (a) shows position of shock varying with angular momentum $\lambda$ and energy $\epsilon$ over time for model D3 $[\gamma = 1.4, \lambda =3.58  \rm [M], \epsilon = 0.0005 ]$, D5 $[\gamma = 1.4, \lambda =3.86  \rm [M] , \epsilon = 0.0001 ]$ and D6 $[\gamma = 1.4, \lambda =3.6  \rm [M] , \epsilon = 0.0001 ]$.  Panel (b) and (c) shows position of shock varying with adiabatic index $\gamma$ over time. Models in panel (b) are C5 $[\gamma = 4/3, \lambda =3.86  \rm [M], \epsilon = 0.0001 ]$, D5 $[\gamma = 1.4, \lambda =3.86  \rm [M], \epsilon = 0.0001 ]$ and E5 $[\gamma = 1.5, \lambda =3.86  \rm [M], \epsilon = 0.0001 ]$ and models in panel (c)  are C6 $[\gamma = 4/3, \lambda =3.6  \rm [M], \epsilon = 0.0001 ]$, D6 $[\gamma = 1.4, \lambda =3.6  \rm [M], \epsilon = 0.0001 ]$ and E6 $[\gamma = 1.5, \lambda =3.6  \rm [M], \epsilon = 0.0001 ]$ }  
     \label{fig:3}
 \end{figure}
Figure [\ref{fig:3}(a)] shows how the shock location is varying with changing value of angular momentum. All curves are plotted for $\gamma = 1.4$. For lower values of $\lambda$, shock position is located closer to the black hole. Also when the matter has more energy, it gets pushed further. That is illustrated 
by the fact, that the location of shock for model D3 ($\lambda=3.58$  [\rm M] and $\epsilon = 0.0005$) is further, but it appears more steady, than shock in model D6 ($\lambda=3.6$  [\rm M], $\epsilon=0.0001$). 
The pattern for shock existence observed here from GR simulation goes as predicted from pseudo-Newtonian simulation in \citet{sukova2015shocks}. It was presented in the article \cite{sukova2015shocks} that for higher $\gamma$, shock formation and oscillation is possible for lower values of $\lambda $ $\sim$ (3.5  [\rm M] - 3.7  [\rm M]) and shock front oscillation can be seen here for model D5 where $\lambda$ = 3.58[M].

In Figure [\ref{fig:3}(b)] and [\ref{fig:3}(c)], the model with $\gamma=4/3$ is plotted with blue line, $\gamma=1.4$ with red line and  $\gamma = 1.5$ is shown with green line.
Figure [\ref{fig:3}(b)] shows here that the lowest adiabatic index leads to the long term oscillations of the shock bubble, the intermediate one to expanding shock and the highest one to faster expansion of the shock.
Figure [\ref{fig:3}(c)] shows the variation of shock position with $\gamma$ for same $\epsilon$ = 0.0001 but with lower $\lambda$ = 3.6 [M].  We can see here, that for the lowest $\gamma$ the shock is accreted,
the intermediate value of adiabatic index leads to quite steady position of the shock, whilst the flow with the highest adiabatic index creates an expanding shock. Similar trend holds also for the three cases with $\lambda = 3.86$ [M], as shown in Figure [\ref{fig:3}(b)].
Hence, the trend seen in our simulations is when other parameters of the gas are kept constant, the flow with higher adiabatic index produces larger shock bubbles. The value of $\gamma$ can even change the nature of the shock behaviour, and increase in the adiabatic index prevents the shock from accretion or blows the shock expansion, while the shock was steady for lower $\gamma$.
This is an important conclusion, because the internal properties and microphysics of the gas are affecting the evolution of the flow as much as the global parameters. Angular momentum and energy of the gas are mainly given by the way, how the gas was supplied to the black hole, e.g. by the properties of the companion star or the temperature of accreting cloud. The adiabatic index corresponds to the internal conditions in the flow, i.e., the degree of ionization and thermo-dynamical properties. In addition, if the adiabatic index would vary along the flow, the shock evolution can further be affected by that.



 \subsection{Dependence of solutions on the adiabatic index}
 \label{sec:dependence}
 We studied models with six different values of $\gamma$: $[1.02, 1.2, 4/3, 1.4,1.5, 5/3]$ with different sets of value for specific angular momentum and specific energy. Models A$(\gamma = 1.02)$, B$(\gamma = 1.2)$, C$(\gamma = 4/3)$, D$(\gamma =1.4)$, E$(\gamma = 1.5)$, F$(\gamma = 5/3)$ and spinning black hole model H $(\gamma= 1.4)$ correspond to the  initial condition where the angular momentum is scaled according to Eqn. (4) (see section 3.2.1 of \citet{sukova2017shocks} for different ways of scaling angular momentum).

For $\gamma = 1.02$, the disk is dominated by the supersonic inflow. 
There is no formation of any outflow and no shock propagating outward for the pairs of ($\lambda,\epsilon$) we have chosen for our study. It is expected that with higher angular rotation, the sonic surface topology changes for this case too, but after much longer time in comparison to other $\gamma$'s.
Nevertheless, accreting flows with such small adiabatic index have very high degrees of freedom. For the chosen parameters in our computation, we have not seen the shock on the equator. The only inference drawn from this model is that the outer sonic point lies very far from the black hole thus the flow is highly supersonic down to the black hole. We checked the positions of critical points with our old pseudo-Newtonian code and we have found out, that for the lower values of chosen $\lambda$'s, the inner sonic point does not even exist, and for higher $\lambda$'s the Rankine-Hugoniot conditions are not satisfied, hence the shock does not exists, which is in agreement with the current GR simulations. 
In previous work by \citet{1987stationary}, it has been pointed out that there exist a unique relation between the opening angle of the shock and the ratio of specific heats, i.e. the adiabatic index of the flow. In our future work, we intend to extend our models for this value of $\gamma$ in 3D as well to see for the existence of non- equatorial shocks.

\begin{figure}
      \textbf{ t = 4000[M]}\\
     \includegraphics[width = 0.5\textwidth]{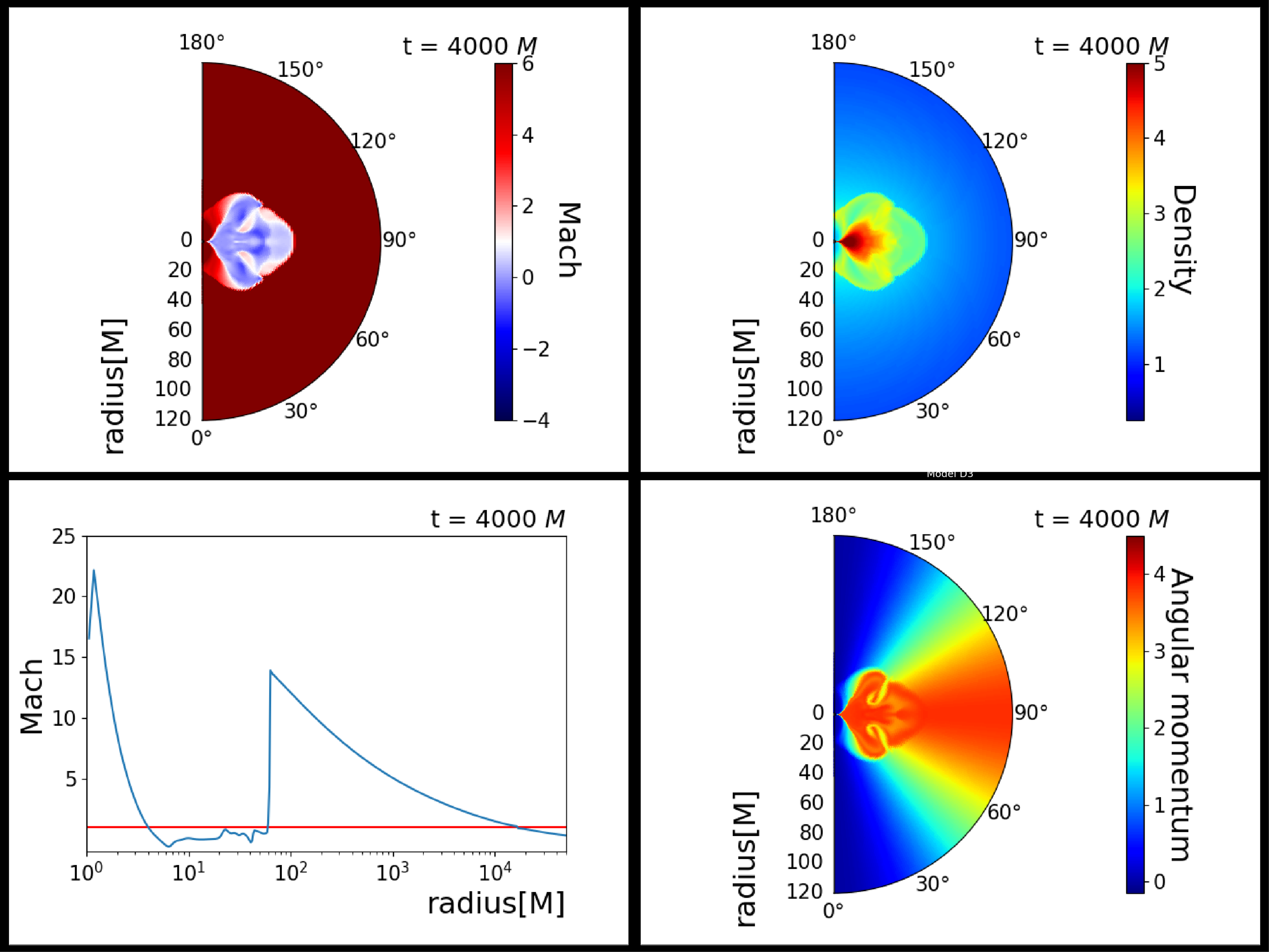}
     \caption{Model B5$[\gamma = 1.2, \lambda =3.86  \rm [M], \epsilon = 0.0001 ]$ at the time  $t = 4000  \rm [M]$ showing the formation of shock bubble. The flow is dominantly supersonic outside the shock front. The equatorial Mach profile shows the high value of Mach close to the black hole horizon. }
     \label{fig:4}
 \end{figure}

In Figure [\ref{fig:4}] we show the formation of shock bubble. The outer sonic point is very far away from the black hole for model B5 which has a small value of $\gamma$ = 1.2. The subsonic region with the shock bubble is present at time t = $ 4000[M] $. There is a very small oscillation at the beginning of the evolution but it gets accreted through inner sonic point  during the further evolution over time. The flow is dominantly supersonic.
The compression ratio $\mathcal{R}$  for model B5 at t = 0[M] is 146.5 which shows very high post shock density. The pressure dominates over rotation and the shock bubble puffs up due to this effect. However, the outer sonic point is located at r$_{c}$=16387[M] which is very far from r$_{s}$ = 27.3[M], so the shock is not able to expand through it. At later times, the shock bubble tries to accrete through the inner sonic point r$_{c}$=3.7[M] as the rotation tries to dominate, but due to considerably high post-shock density even at t = 2000[M] we have $\mathcal{R}$ = 5.45. Then, the shock bubble takes an interesting shape. As can be seen in Figure [\ref{fig:4}], the filament kind of structures develop in the shock bubble. As the flow evolves, shock starts oscillating for a while and then gets accreted through the inner sonic point. 

\begin{figure}
     \includegraphics[width = 0.5\textwidth]{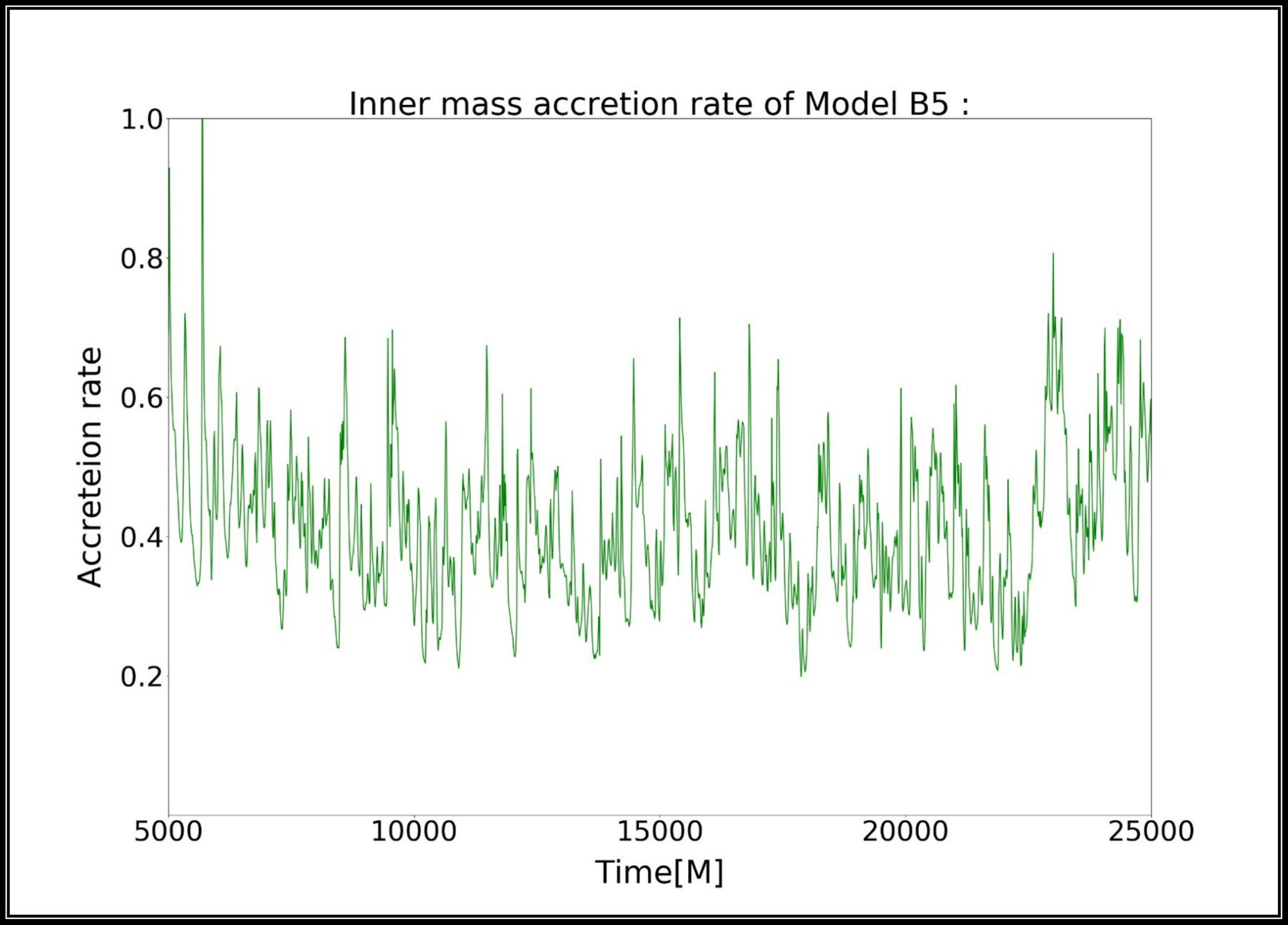}\\
     \caption{The inner mass accretion rate for Model B5 $[\gamma = 1.2, \lambda =3.86  \rm [M] , \epsilon = 0.0001 ]$ is shown here. The end time of the simulation, t = $10^{6} \rm [M] $. See Section \ref{sec:frequency} and Table [\ref{Table 2}] for details. }
     \label{fig:5}
 \end{figure}

\begin{figure}
     \textbf{(a)}\\
       \includegraphics[width = 0.5\textwidth]{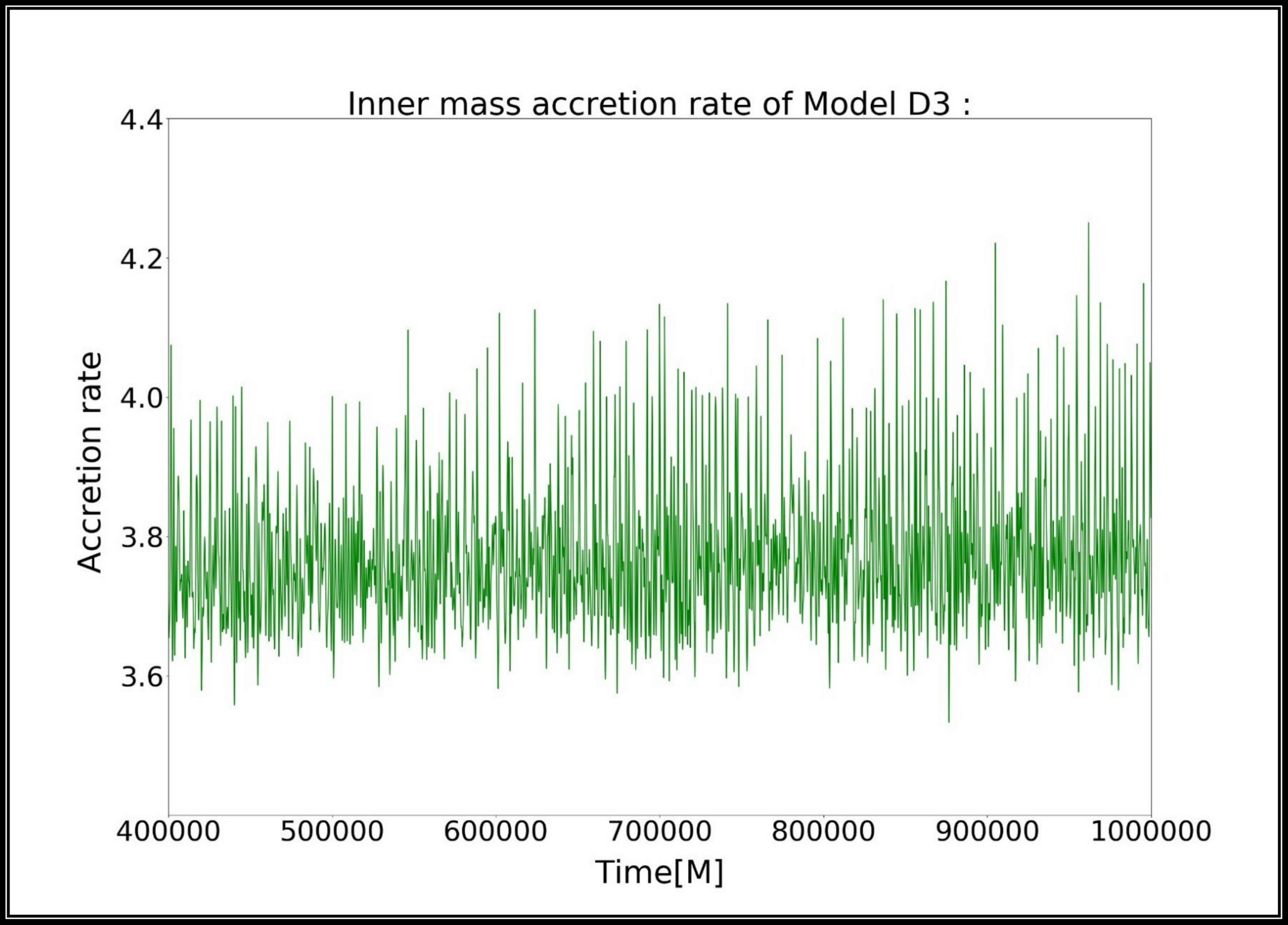}
     \textbf{(b)}\\
      \includegraphics[width = 0.5\textwidth]{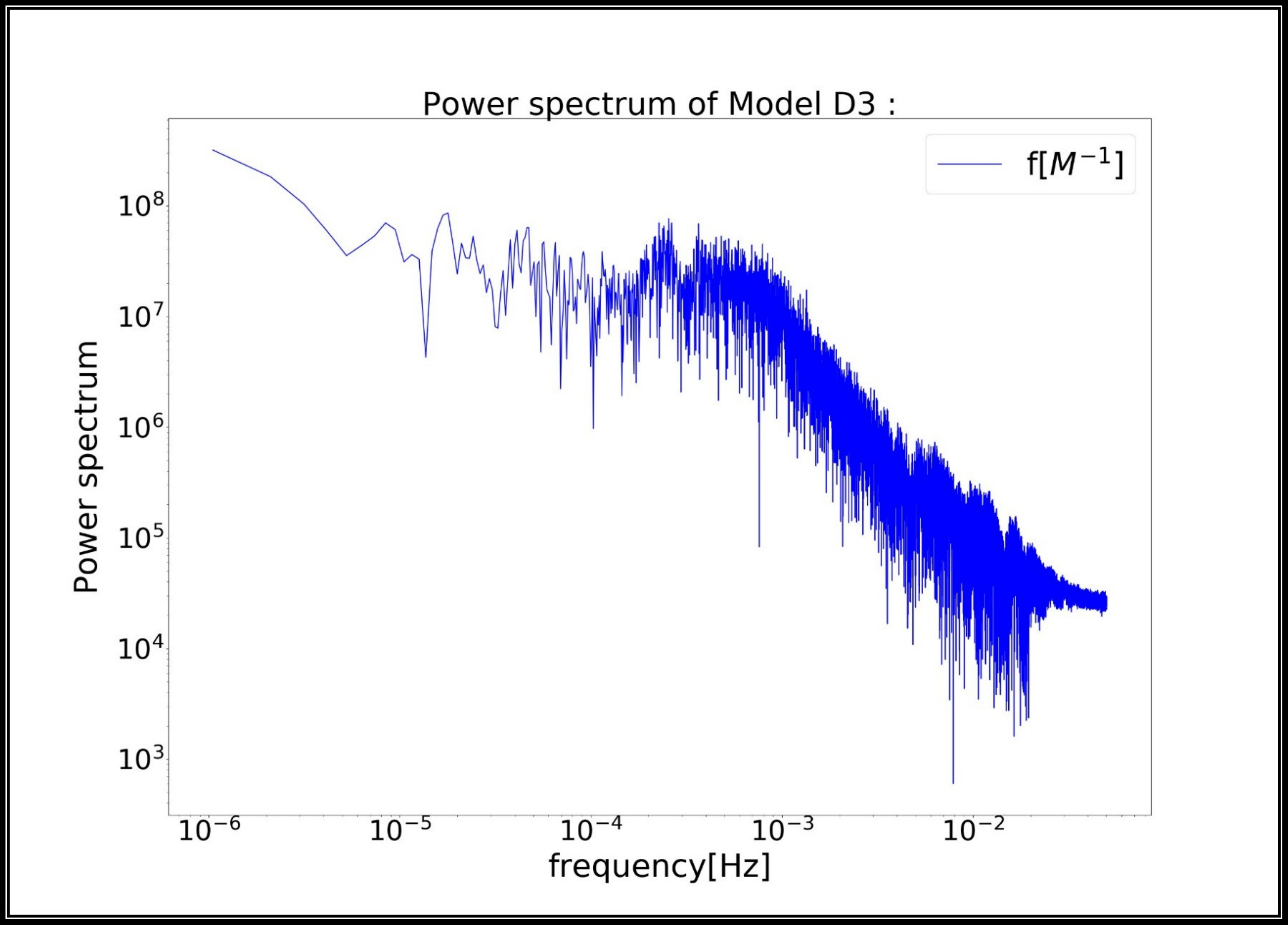}
    \caption{The inner mass accretion rate for Model D3 $[\gamma = 1.4, \lambda =3.58  \rm [M] , \epsilon = 0.0005 ]$ and power density spectrum has been shown here. Two small peaks can be seen in the PDS. See Section \ref{sec:frequency} and the Table [\ref{Table 2}] for values of frequency. }
    \label{fig:6}
\end{figure}

 Figure [\ref{fig:5}]  and Figure [\ref{fig:6}(a)] show the variability of the mass accretion rate with time for model B5 and D3 respectively. (See the x scale showing the shorter range of time for which shock oscillation exists in model B5 in Figure [\ref{fig:5}]).
 With $\gamma = 1.2$ only model B5 (specific angular momentum provided as 3.86[M] and very small specific energy as 0.0001) shows small oscillation. Other models with lower values of angular momentum do not show any oscillation and forms only one sonic point.  The fact here is that for low $\gamma$, the sound speed is low compared to flow velocity and the radius of the initial outer sonic surface is much farther for other values of angular momentum $\lambda <$ 3.86[M].

  \begin{figure}
     \textbf{(a) t = 4000[M]}\\
     \includegraphics[width = 0.5\textwidth]{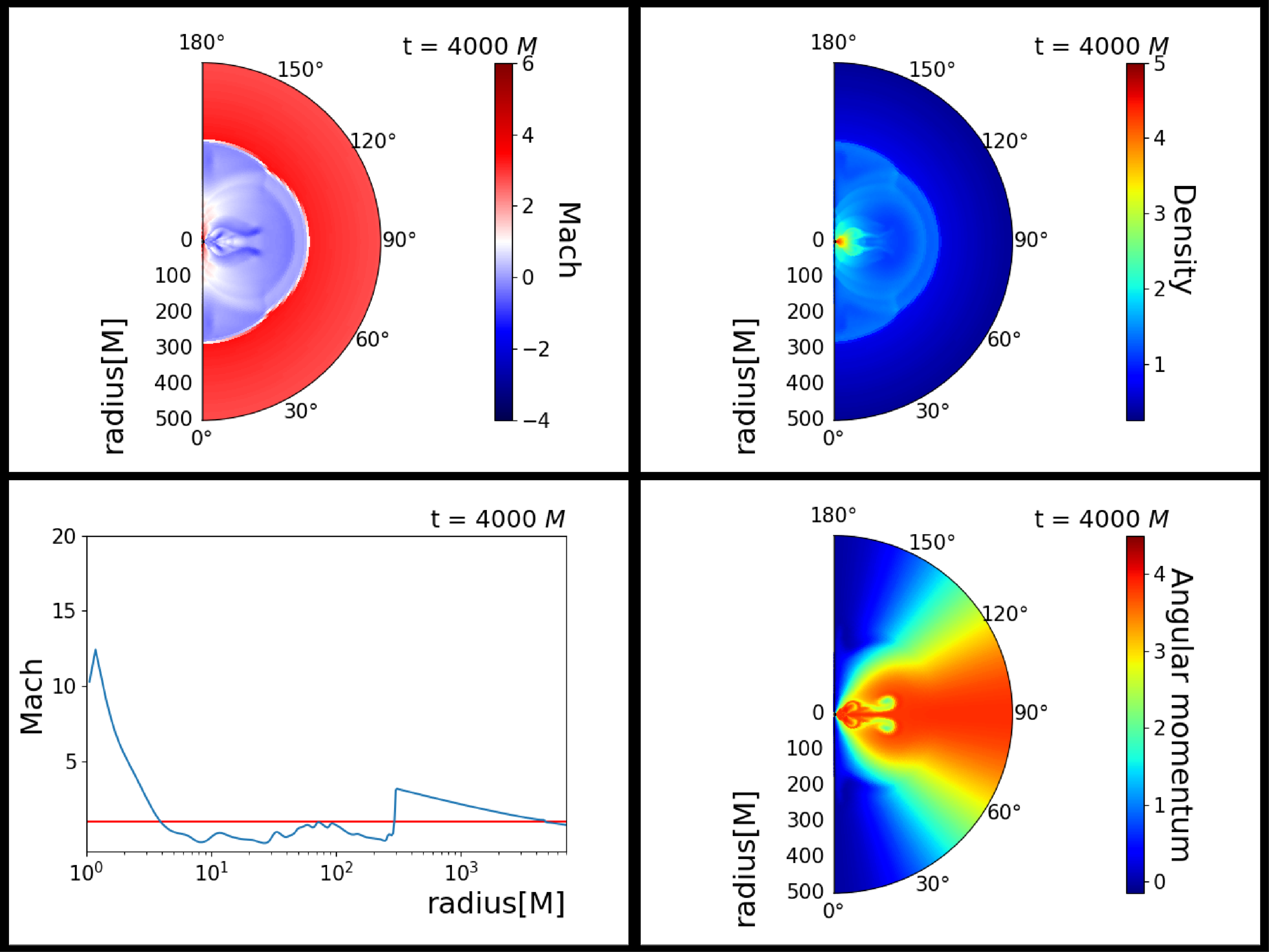}
     \textbf{(b) t = 86000[M]}\\
     \includegraphics[width = 0.5\textwidth]{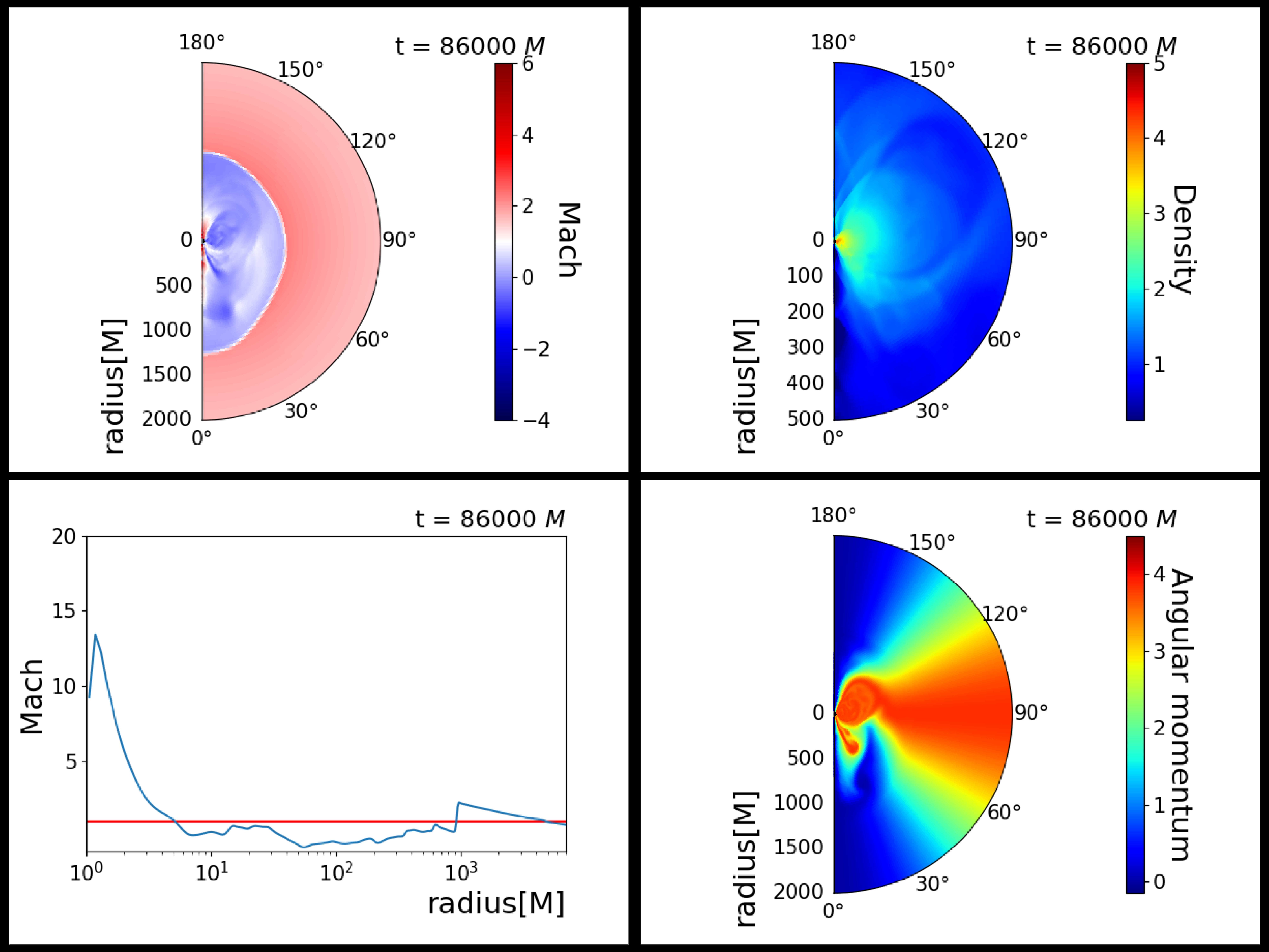}
     \textbf{(c) t = 326000[M]}\\
     \includegraphics[width = 0.5\textwidth]{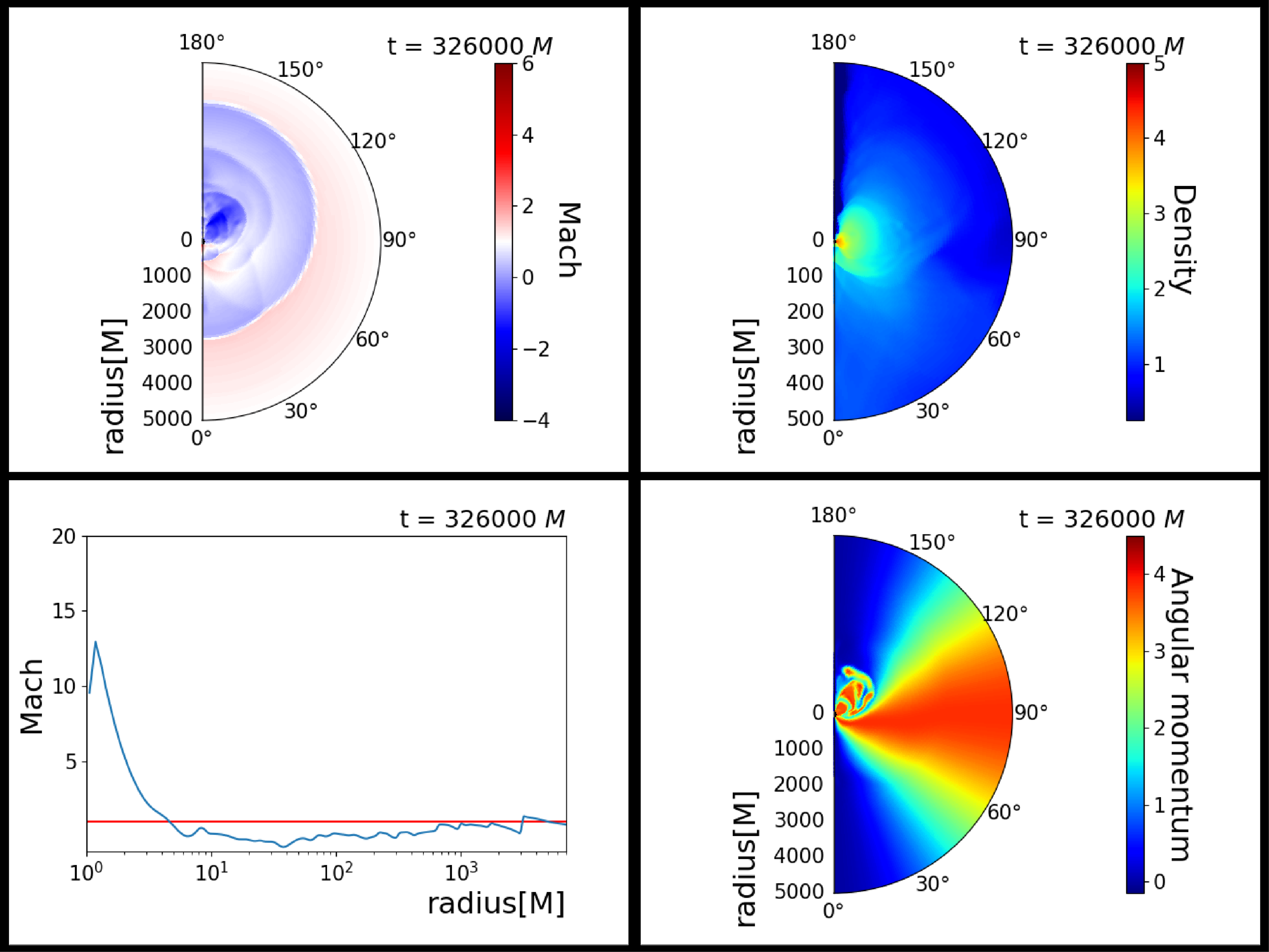}
     \caption{Model D5 [$\gamma$ = 1.4, $\lambda$ = 3.86 [\rm  M], $\epsilon$  = 0.0001] at t = 4$^{3}$ [M],  t = 8.6$*10^{4}$ [M] and  t = 3.26$*10^{5}$ [M] showing the Mach number, angular momentum and density profile. Small oscillation occurs at the beginning of the evolution but later it has been seen subsonic flow dominated over the upper half. Flow is highly asymmetric.}
     \label{fig:7}
 \end{figure}

Figure [\ref{fig:7}] shows model D5  [$\gamma$ = 1.4, $\lambda$ = 3.86 [M], $\epsilon$  = 0.0001] at three time snapshots. Different size of the shock bubble can be compared for models B5 and D5 (cf. Figures [\ref{fig:4}] and [\ref{fig:7}(a)]). The different scale for radius in Figures [\ref{fig:7}(a)], [\ref{fig:7}(b)], [\ref{fig:7}(c)] shows the growth of shock bubble to a very large radius, which is about six times larger in model D5 than in B5. At later time during the evolution of the flow the subsonic bubble has moved northwards and the flow lost the symmetry with respect to the equatorial plane. The angular momentum profile shows some mixing of low and high angular momentum close to the black hole, and the shock is oscillating also in vertical direction.
We intend to investigate this asymmetry of the flow in more detail in the future 3D simulation.

\begin{figure}
       \textbf{(a) t = 208000[M]}\\
     \includegraphics[width = 0.5\textwidth]{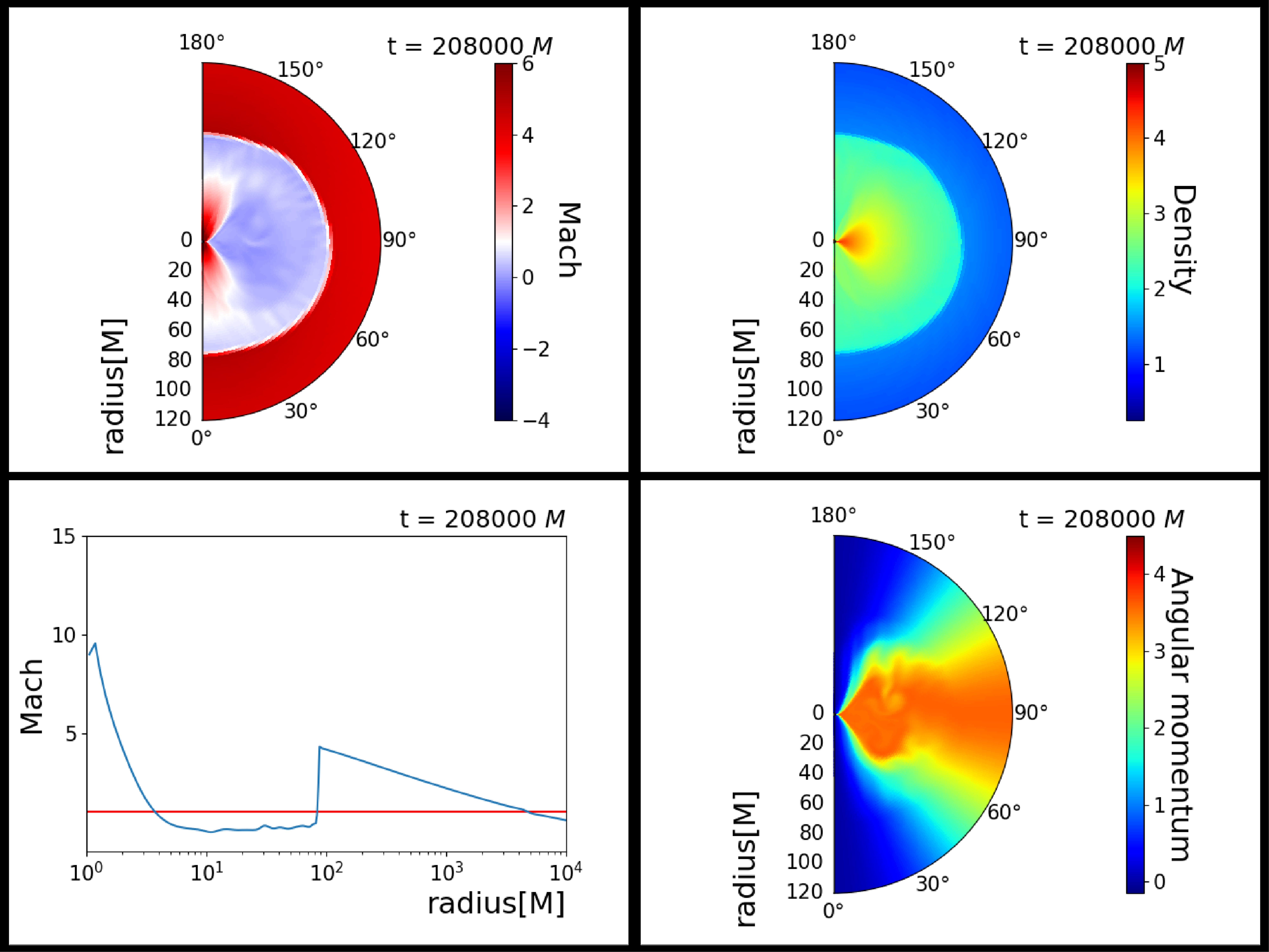}
     \textbf{(b) t = 496254[M]}\\
     \includegraphics[width = 0.5\textwidth]{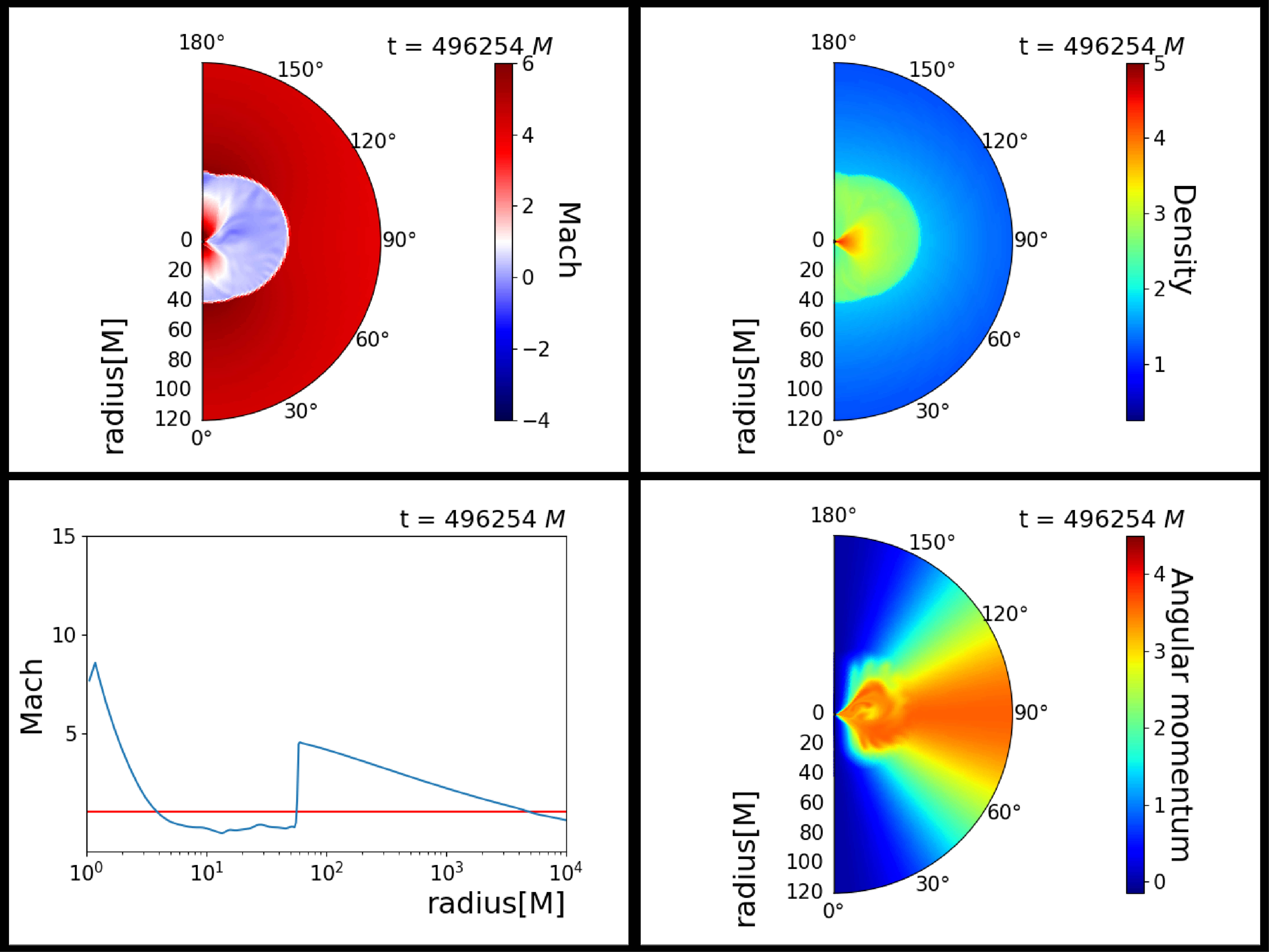}
     \textbf{(c) t = 879850[M]}\\
     \includegraphics[width = 0.5\textwidth]{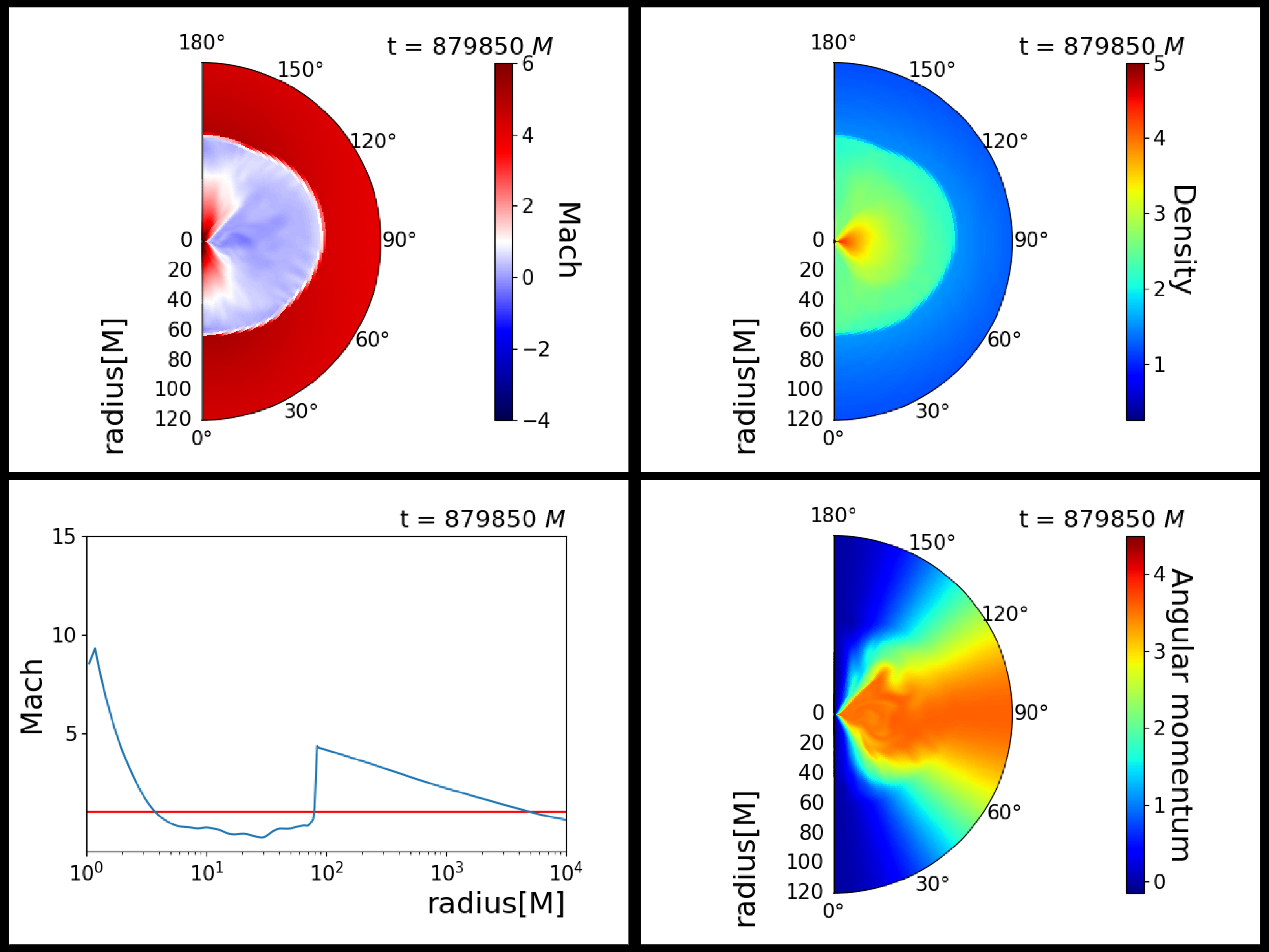}
     \caption{Model D6 $[\gamma = 1.4, \lambda =3.6  \rm [M], \epsilon = 0.0001 ]$ at three snapshots of evolution of the flow showing the Mach number, angular momentum and density profile. The oscillation of the shock front can be seen in all the profiles. See text and table for details.}
     \label{fig:8}
 \end{figure}

Figure [\ref{fig:8}] shows on several snapshots how the shock position oscillates in time for model D6 $[\gamma = 1.4, \lambda =3.6  \rm [M], \epsilon = 0.0001 ]$ as the gas pressure keeps pushing the shock front back and forth. The amount of pressure created working against shock depends on the values of $\gamma$  and it can be seen that much less pressure is created for this adiabatic index compared to model E6.
The shock keeps oscillating between $r_{s}=$ 44.4[M] and 97.5[M]. The compression ratio $\mathcal{R}$, i.e the ratio of post shock density to that of pre-shock  at the maximal and minimal shock position close to the black hole comes as   $\mathcal{R}=3.04$  and  $\mathcal{R}=2.49$ correspondingly.


   \begin{figure}
     \textbf{(a) t = 26000[M]}\\
      \includegraphics[width = 0.5\textwidth]{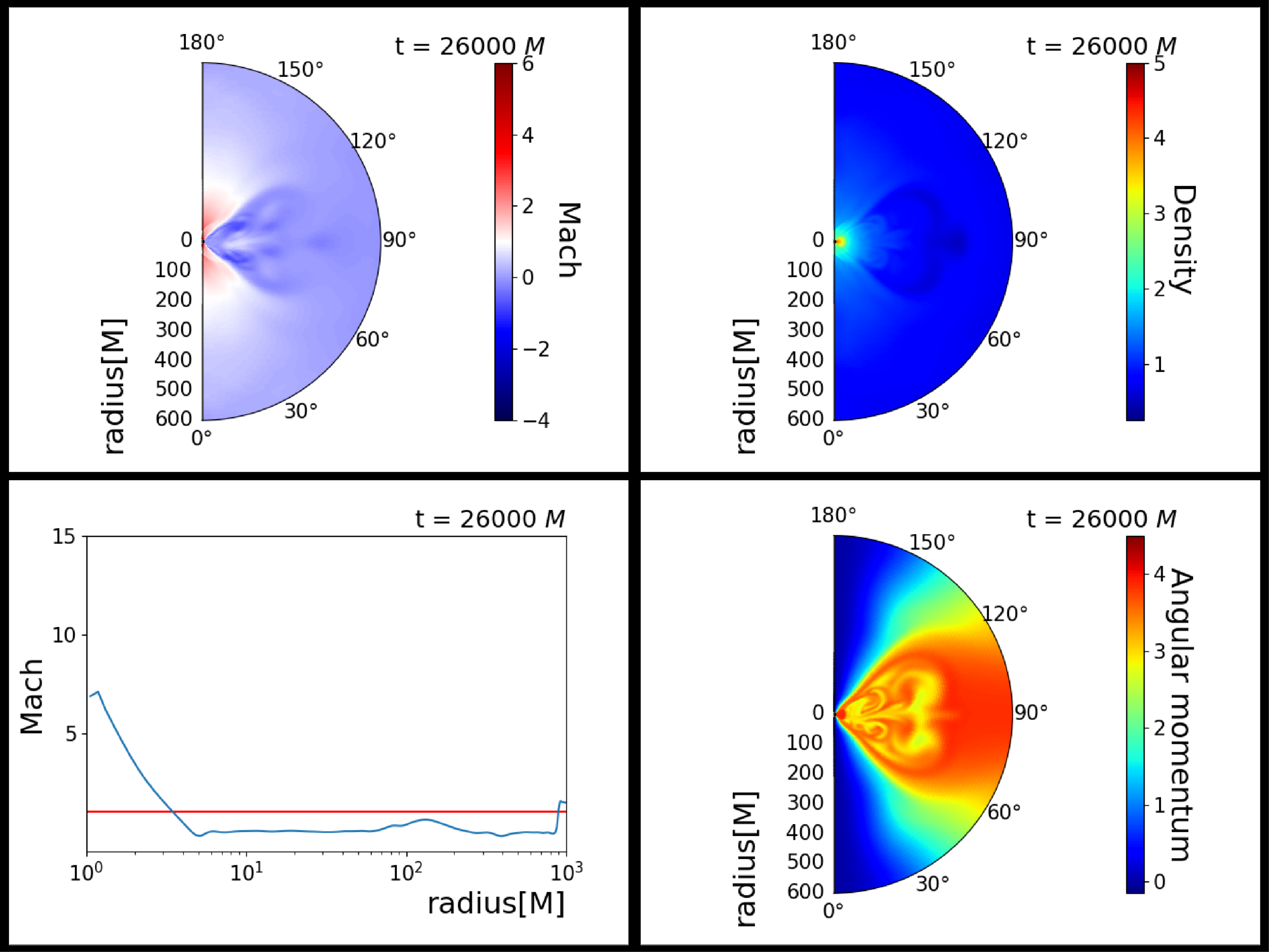}
     \textbf{(b)}\\
     \includegraphics[width = 0.5\textwidth]{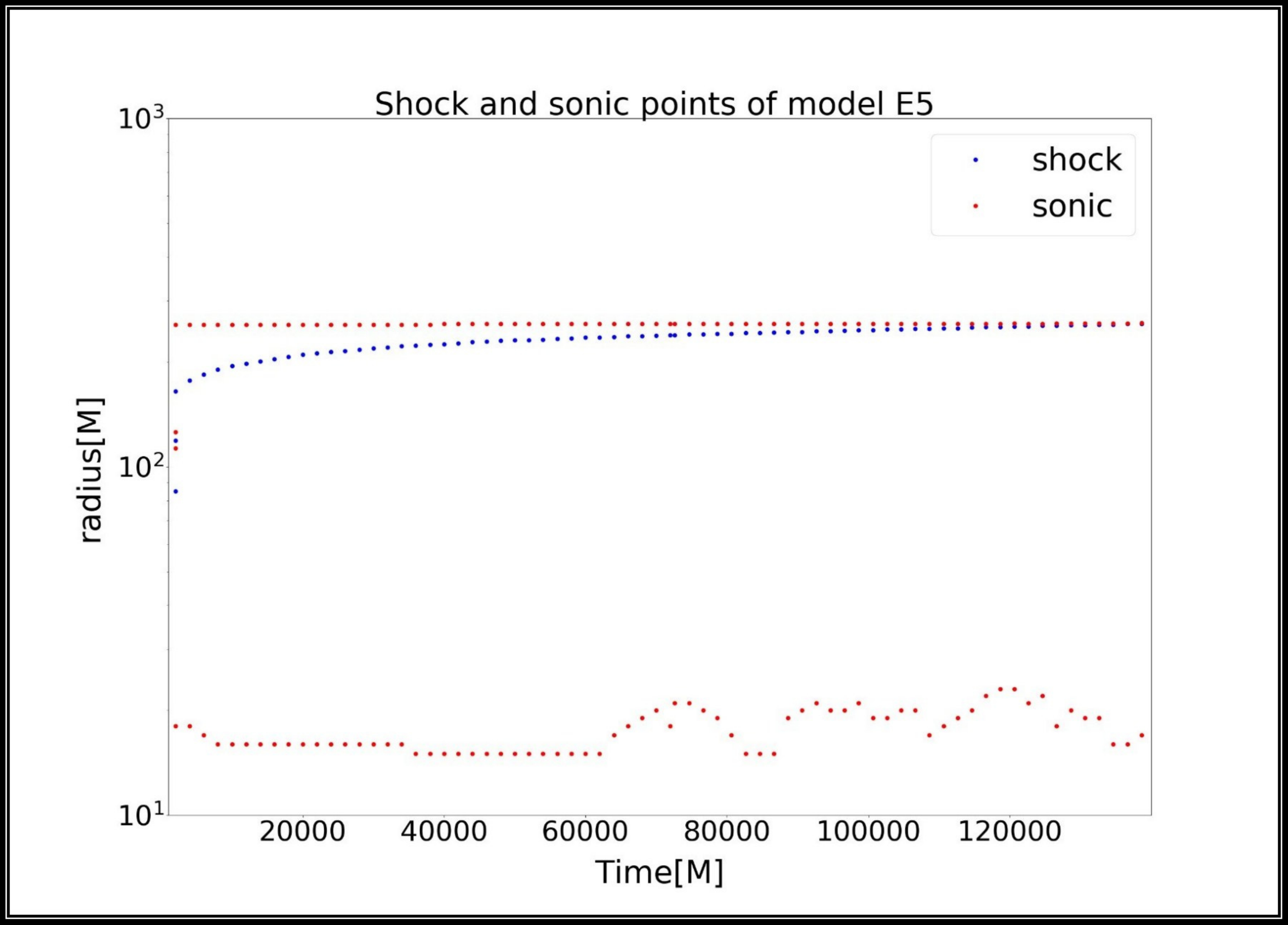}
     \caption{Model E5 $[\gamma = 1.5, \lambda =3.86  \rm [M], \epsilon = 0.0001 ]$ at t = $26*10^{3}$  \rm [M] showing the mach ,angular momentum and density profile  in the upper half and bottom part shows the expansion of shock position through outer sonic front.}
     \label{fig:9}
 \end{figure}
\begin{figure}
     \textbf{(a) t = 18000[M]}\\
      \includegraphics[width = 0.5\textwidth]{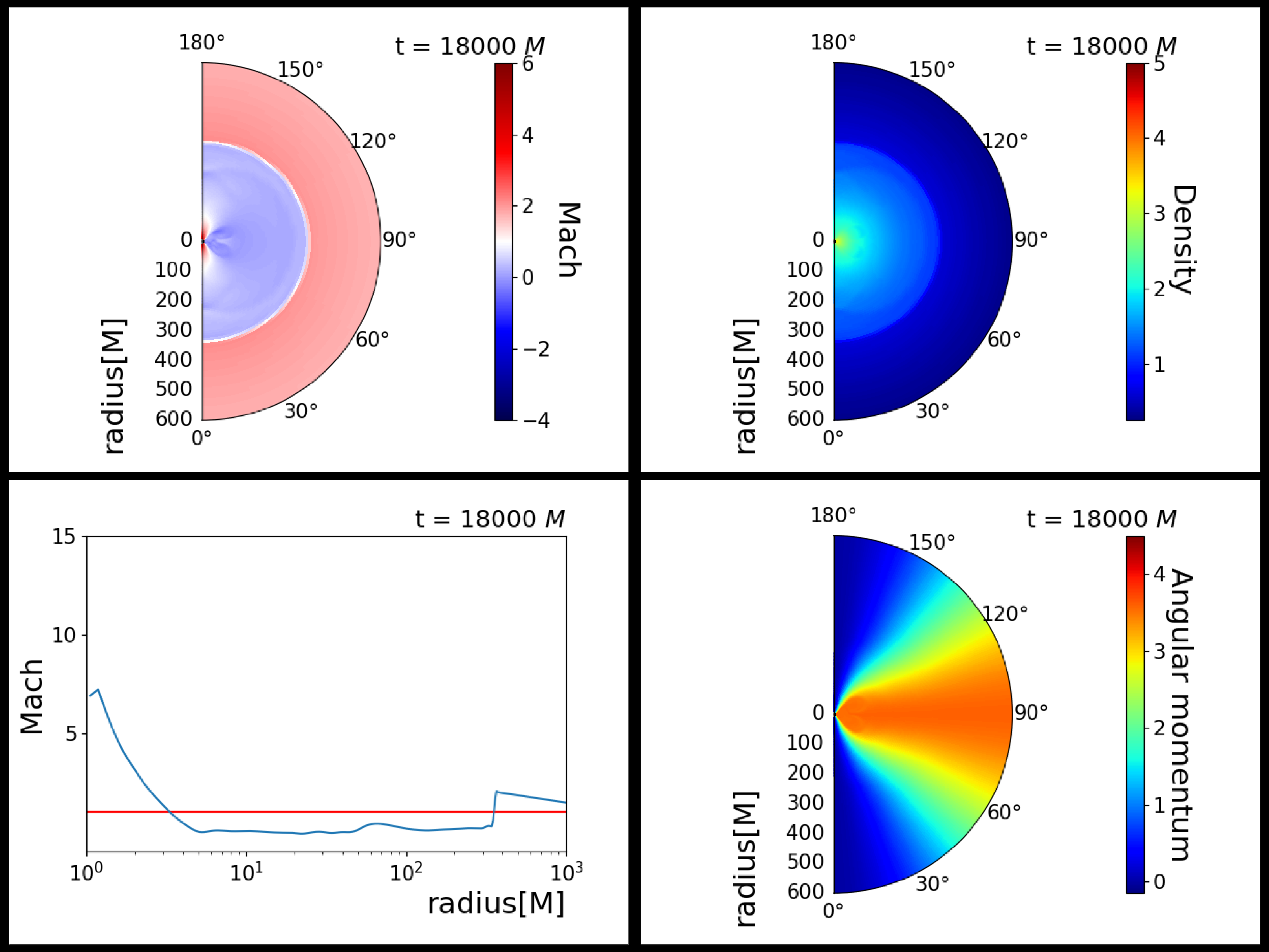}\\
     \textbf{(b) t = 186601[M]}\\
     \includegraphics[width = 0.5\textwidth]{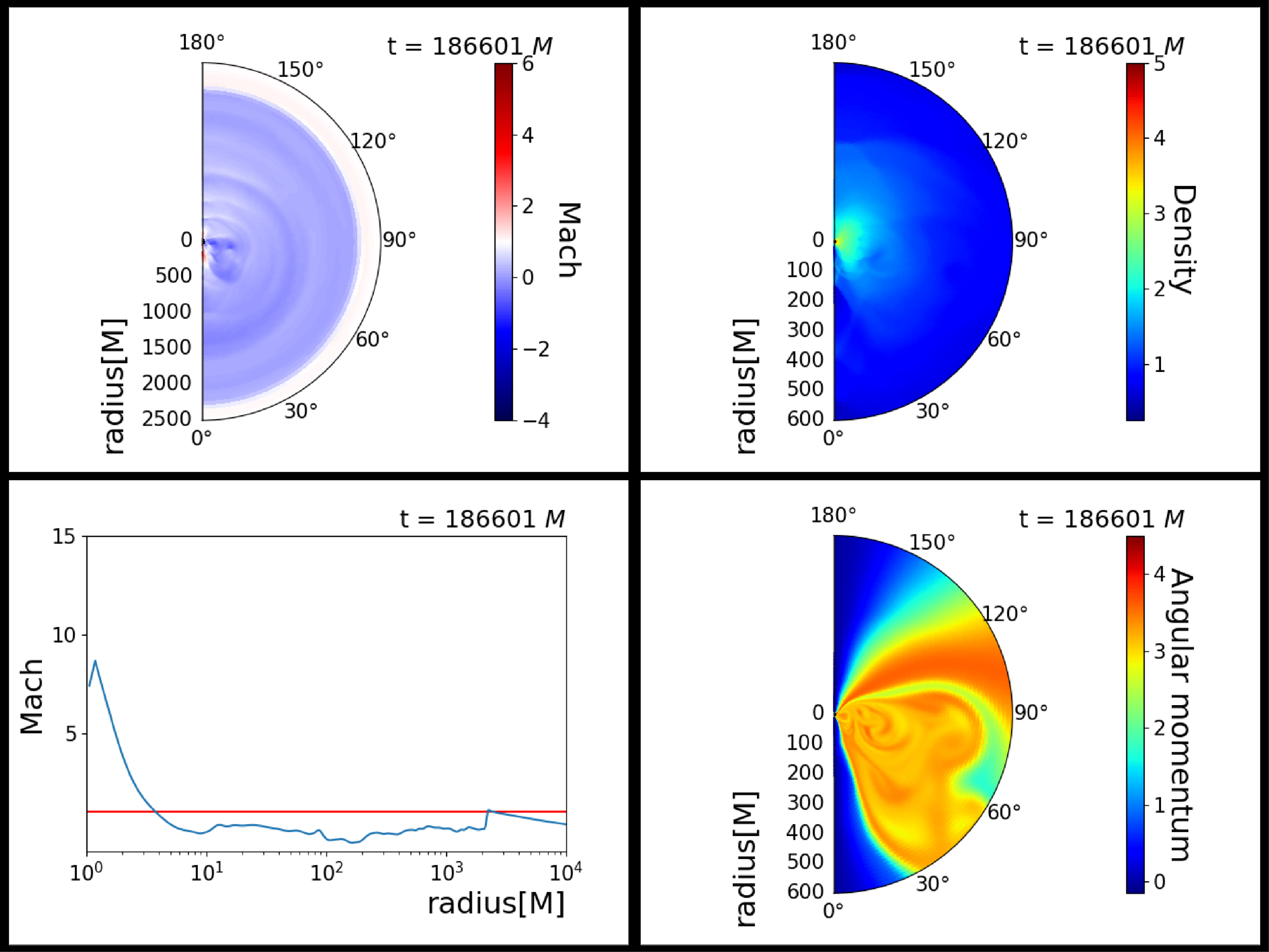}\\
     \caption{Model E6 $[\gamma = 1.5, \lambda =3.6  \rm [M], \epsilon = 0.0001 ]$ at  t = $6000  \rm [M] $ and  t = $178601[M] $ correspondingly. The flow becomes completely subsonic at later times and the shock front expands up to the outer sonic position and disappears. See the different range of radius shown in the two parts for the Mach number, density and angular momentum profile.}
     \label{fig:10}
 \end{figure}

In Figure [\ref{fig:9}], we present model E5  $[\gamma = 1.5, \lambda =3.86  \rm [M], \epsilon = 0.0001 ]$.  For high angular momentum of $\lambda=3.86$M, the shock front expands towards the outer sonic point. With the same amount of specific energy and adiabatic index but with lower angular momentum, model E6 $[\gamma = 1.5, \lambda =3.6  \rm [M], \epsilon = 0.0001 ]$ presented in Figure [\ref{fig:10}] also shows expansion of sonic front. This implies that to have more gas pressure to make the shock oscillate we need to provide very low angular momentum ($<$ 3.5[M] ) to the flow having $ \gamma= 1.5$. In the article \cite{sukova2015shocks}, it has been shown that the shock exists for $\lambda$ between about 3.3  [\rm M] and 3.5  [\rm M] for the same $\gamma$ and $\epsilon$. 

In Figure [\ref{fig:10}], the shock front can be seen expanding up to the outer sonic position for $\gamma$ = 1.5 as the flow evolves with time (see the scale for radius in the Figure [\ref{fig:10}], where the dominant subsonic region reaches to very large radial distances). We checked that the outer sonic point for model E6 is located at 2344 {\rm [M]}.
Contrary to this, flow for $\gamma$ = 1.2 as in model B6 with the same $\lambda$, is highly supersonic. Here the outer sonic point position is farther from the black hole and the shock front is close to the black hole, easily accretes through inner sonic point. 
In order to understand this different behaviour of shock we compare model B6 with E6. The outer sonic point is at 16387 {\rm [M]} for model B6. The gas pressure is higher in the post shock region for model B6 than that for model E6 but the compression ratio for E6 is 10.8 which push the shock out through the outer sonic point. But for model B6, the compression ratio is 7.5 which is not sufficient for the expansion of shock. The structure of the flow is very non-uniform because of the mixing of low and high angular momentum gas in the shock bubble, causing the turbulence in the flow.

\begin{figure}
     \textbf{(a) t = 478325[M]}\\
      \includegraphics[width = 0.5\textwidth]{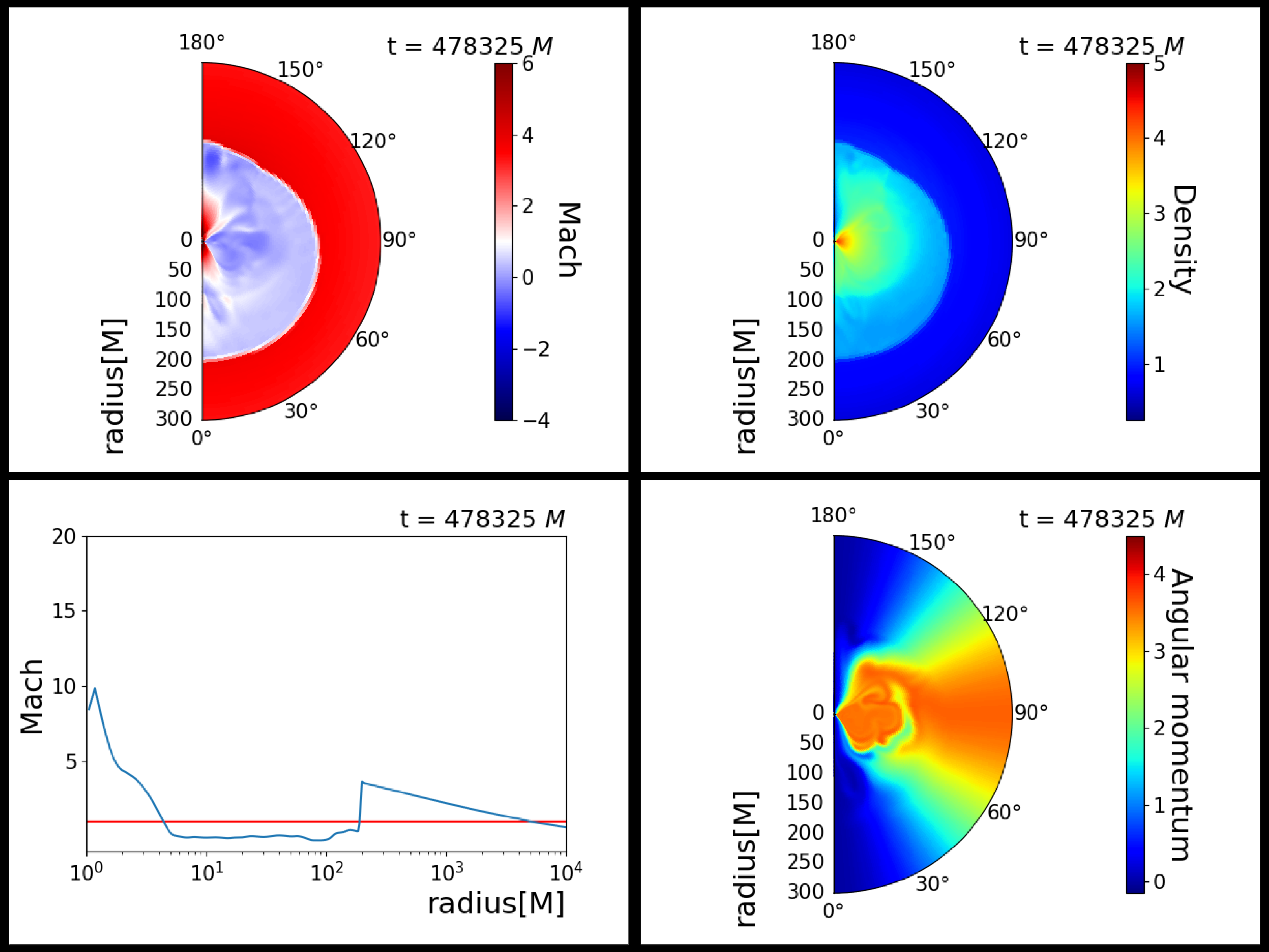}\\
     \textbf{(b) t = 508325[M]}\\
     \includegraphics[width = 0.5\textwidth]{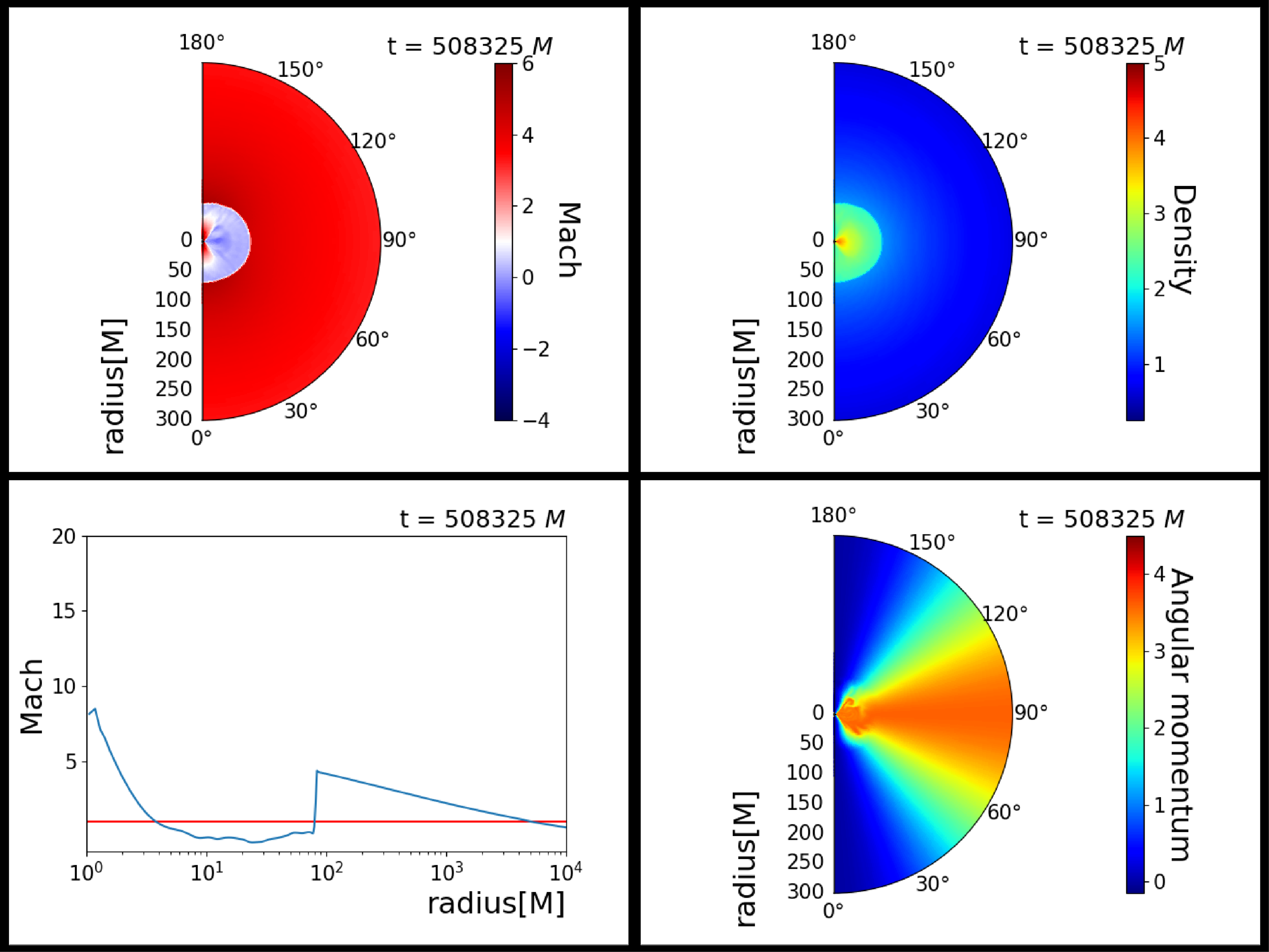}\\
        \textbf{(c) t = 532325[M]}\\
     \includegraphics[width = 0.5\textwidth]{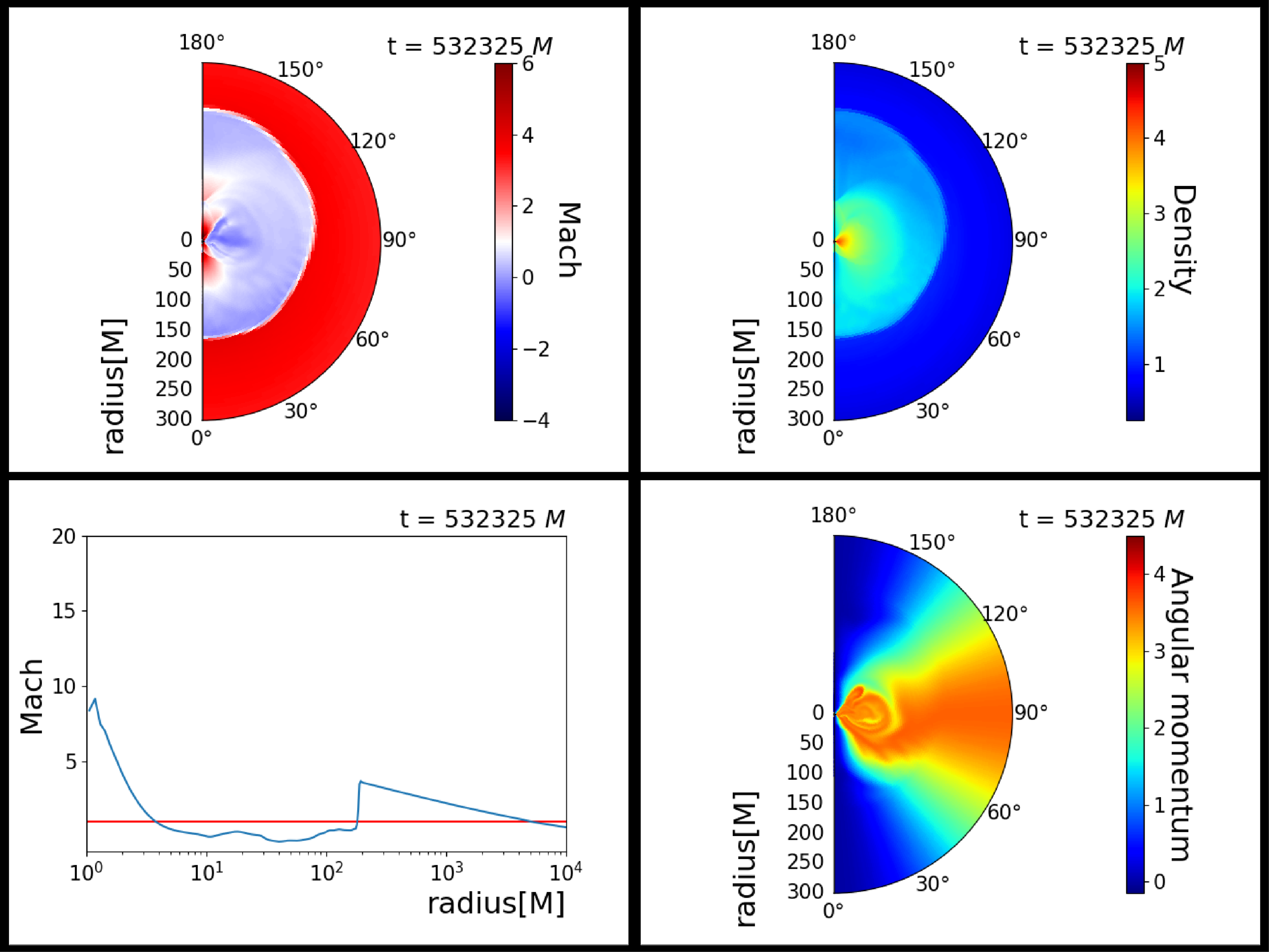}
     \caption{Model H4 $[\gamma = 1.4, \lambda =3.6  \rm [M], \epsilon = 0.0001 ]$ with spin, a = 0.10 at  t = 478325[M], t = 508325[M] and t = 532325[M] correspondingly. Panel (a),(b),(c) show the flow of mass with low angular momentum through equator, shrinking of the shock bubble and then rebuilding of the bubble.}
     \label{fig:11}
 \end{figure}
 
Figure [\ref{fig:11}] shows oscillating shock bubble shrinking in panel (b) and then rebuilding again in panel (c) during the evolution of the flow of model H4 (spinning model, a = 0.10). This interesting occurrence can be interpreted as result of two processes explained in detail in section \ref{sec:frequency}.

\begin{figure}
     \textbf{(a)}\\
      \includegraphics[width = 0.5\textwidth]{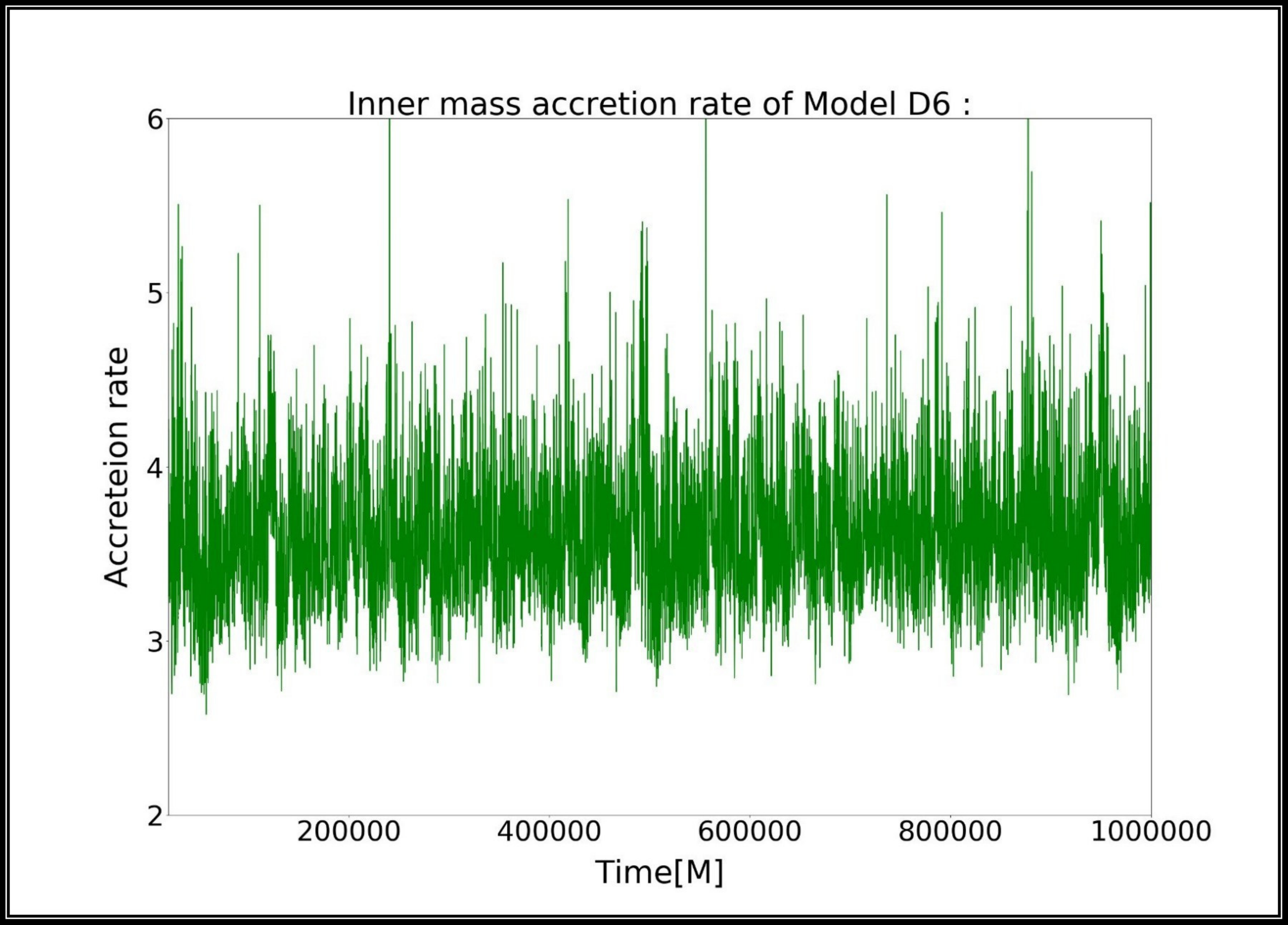}
     \textbf{(b)}\\
     \includegraphics[width = 0.5\textwidth]{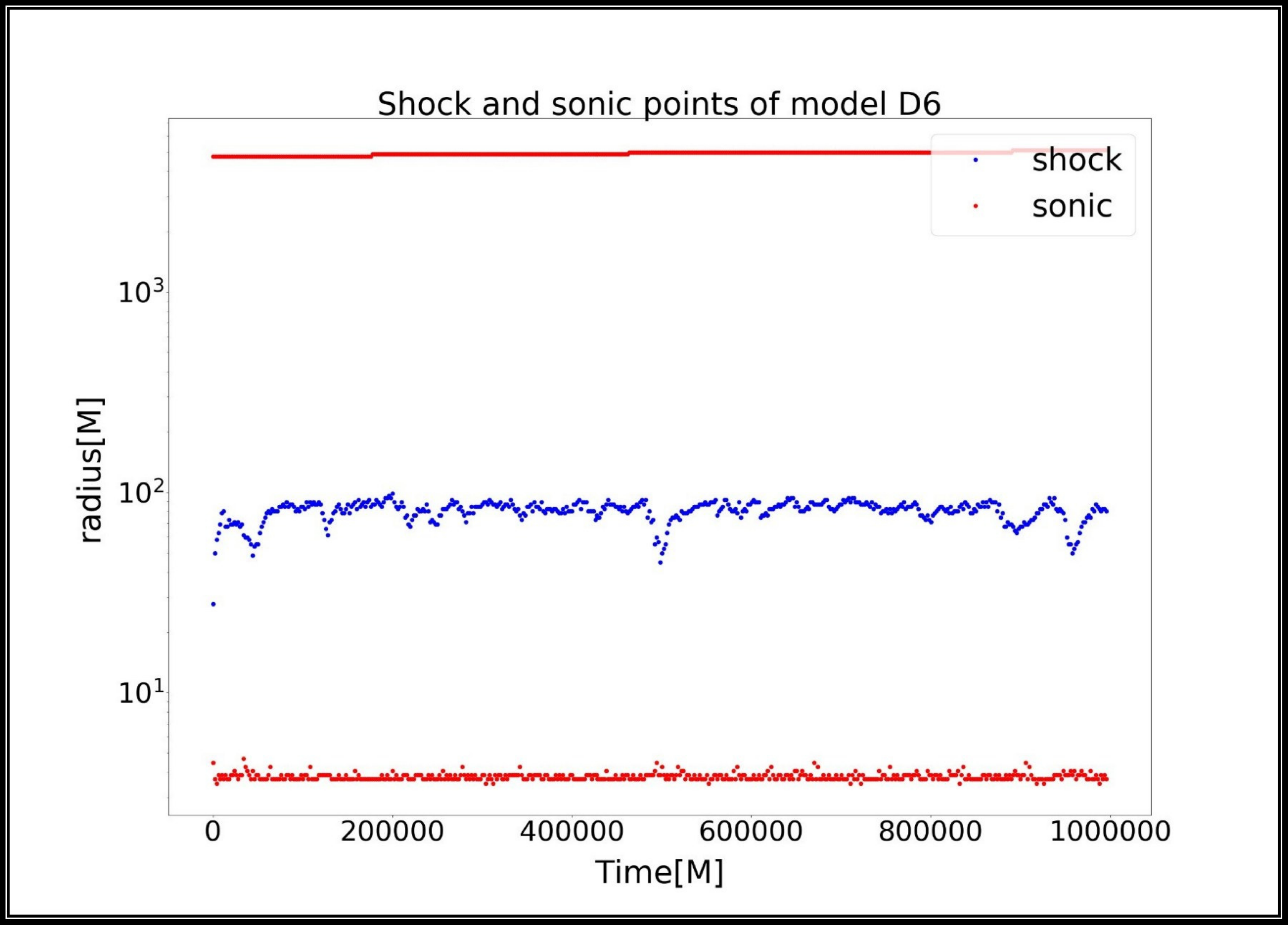}
     \textbf{(c)}\\
     \includegraphics[width = 0.5\textwidth]{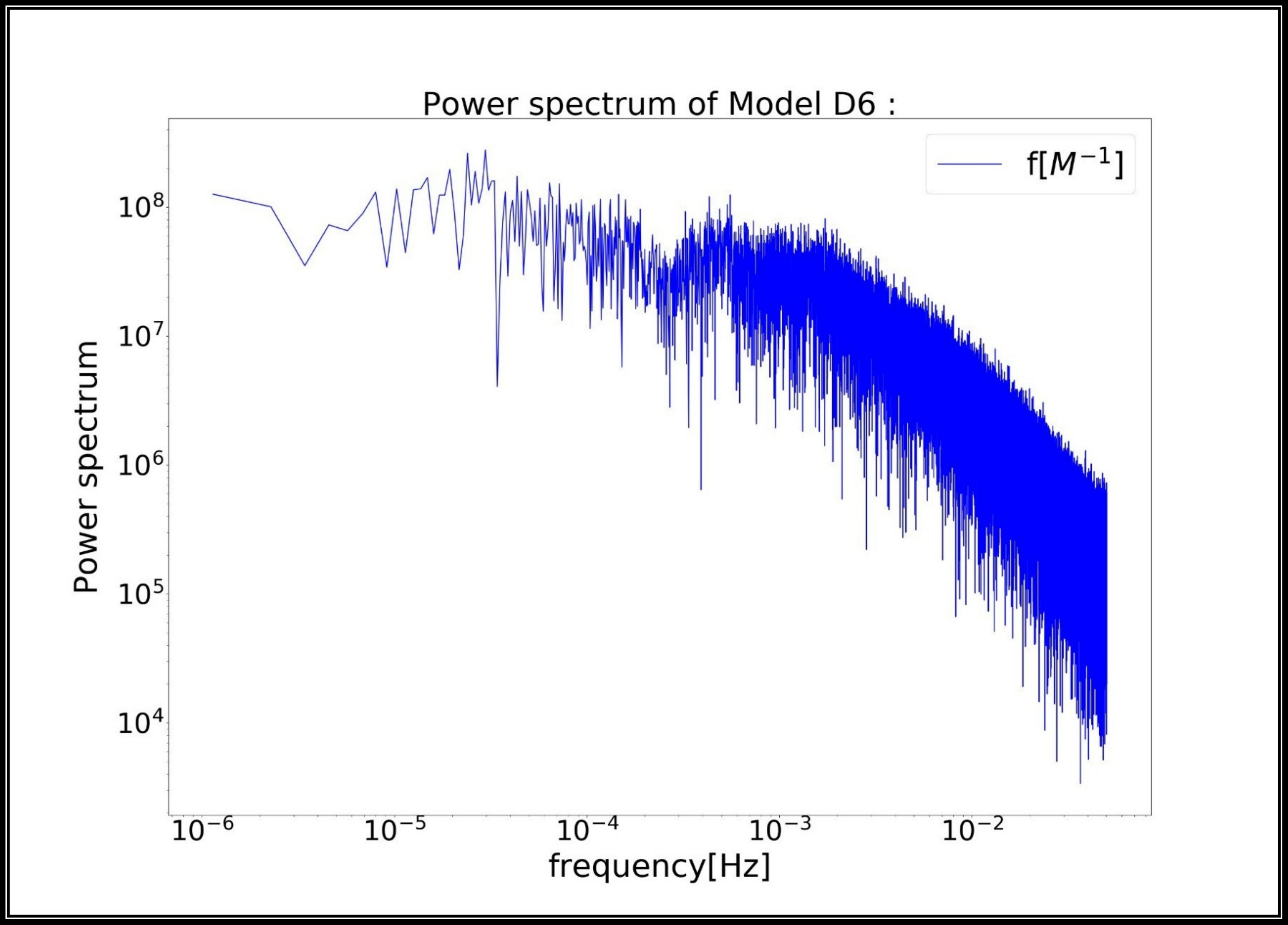}
     \caption{Model D6 $[\gamma = 1.4, \lambda =3.6  \rm [M], \epsilon = 0.0001 ]$. The first figure, 12(a) shows the mass accretion rate followed by the oscillation of shock position over the time evolution of the flow in 12(b). Figure 12(c) shows the power spectrum calculated from the above shown accretion rate. See the values of frequency and other details in the text.  }
     \label{fig:12}
  \end{figure}

To summarize our findings, we present Figure [\ref{fig:12}] 
with the time evolution for the models with $\gamma = 1.4$ 
and show the oscillations of shock fronts for angular momentum, $\lambda=3.6$ [M]. The shock oscillates for a very long time unlike model B5. (See the different timescale for variation in mass accretion rate in Figure [\ref{fig:5}] and [\ref{fig:12}(a))].

\begin{figure}
      \textbf{(a)}\\
      \includegraphics[width = 0.5\textwidth]{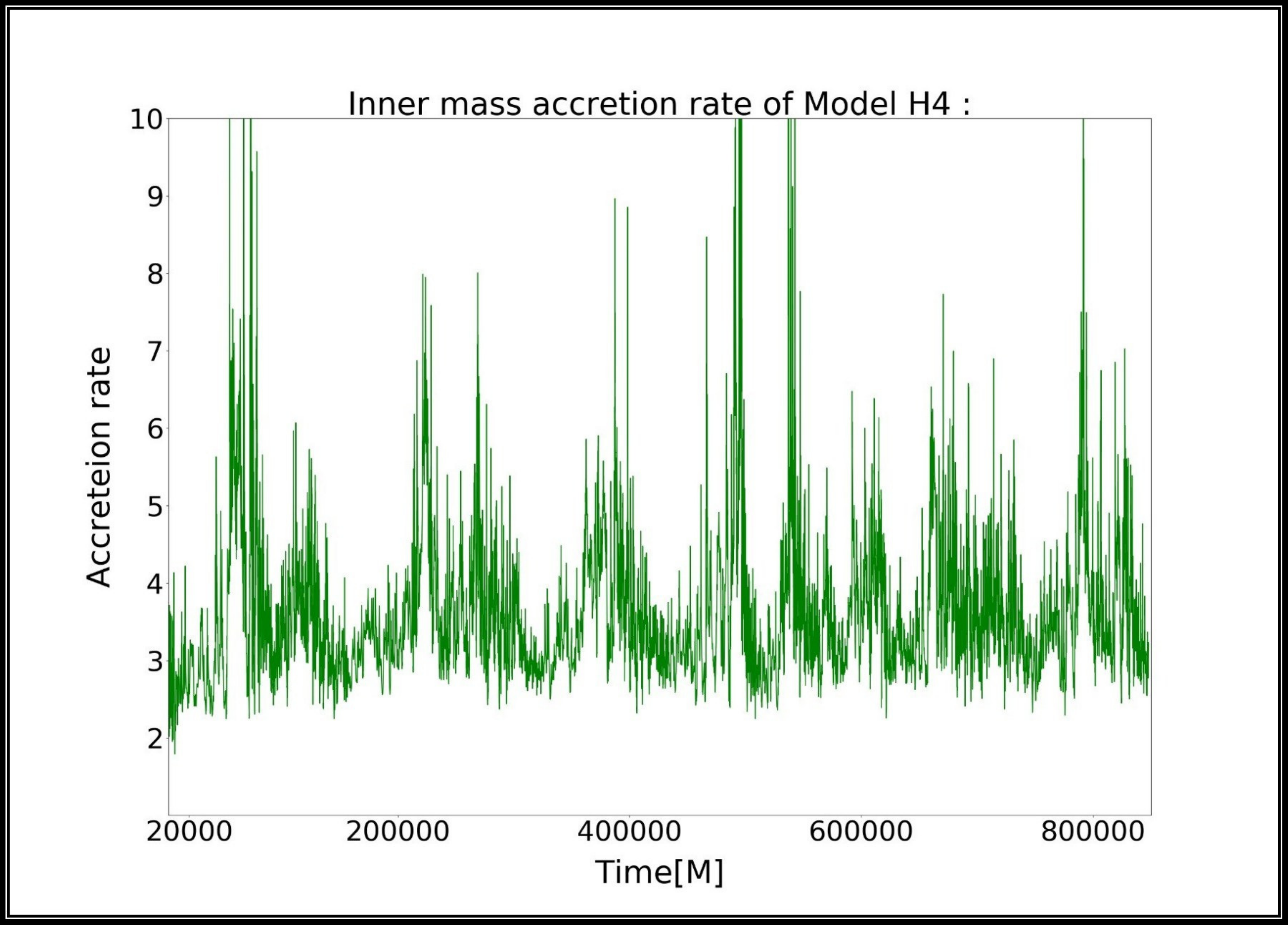}
      \textbf{(b)}\\
     \includegraphics[width = 0.5\textwidth]{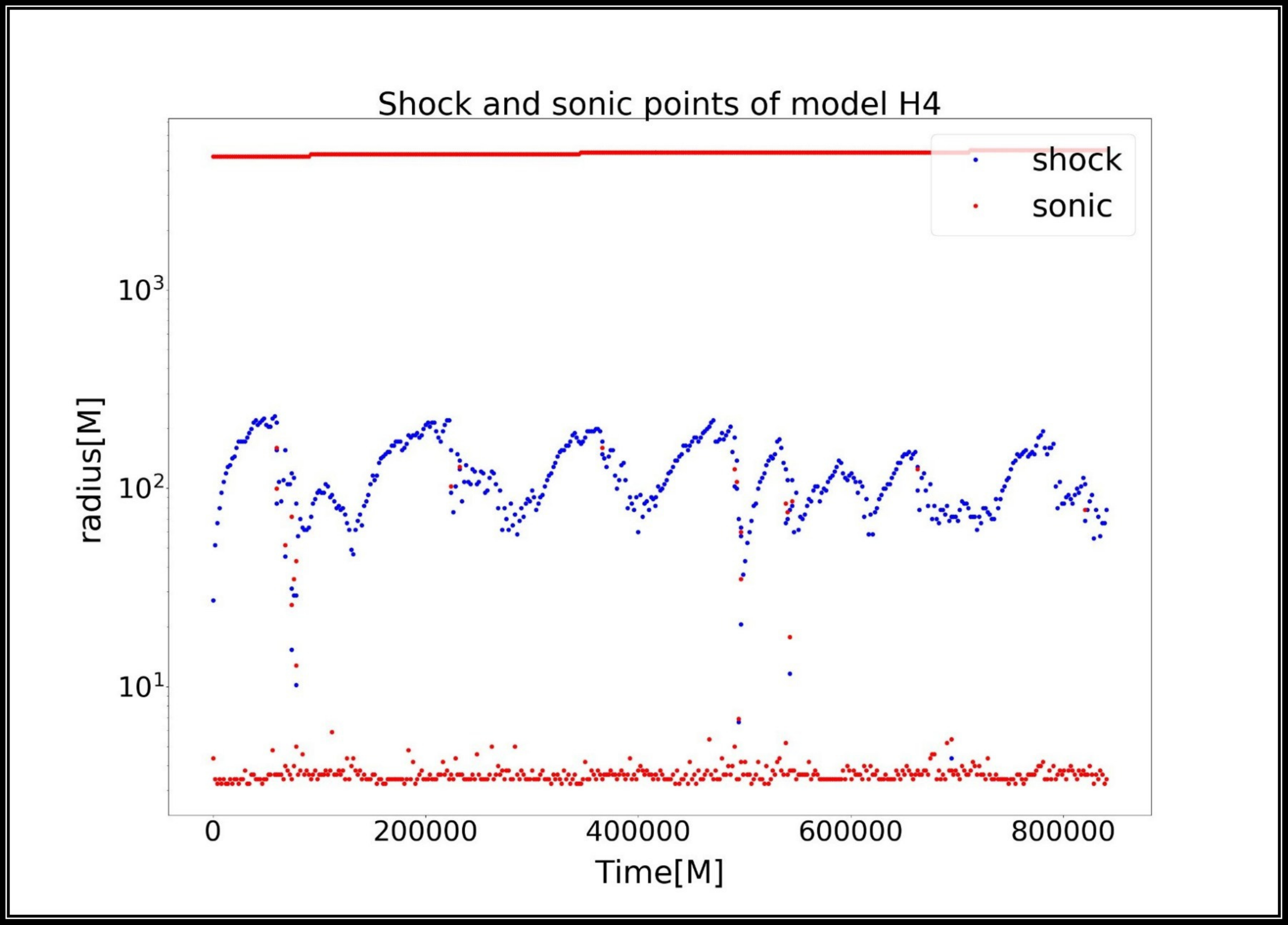}
     \textbf{(c)}\\
     \includegraphics[width = 0.5\textwidth]{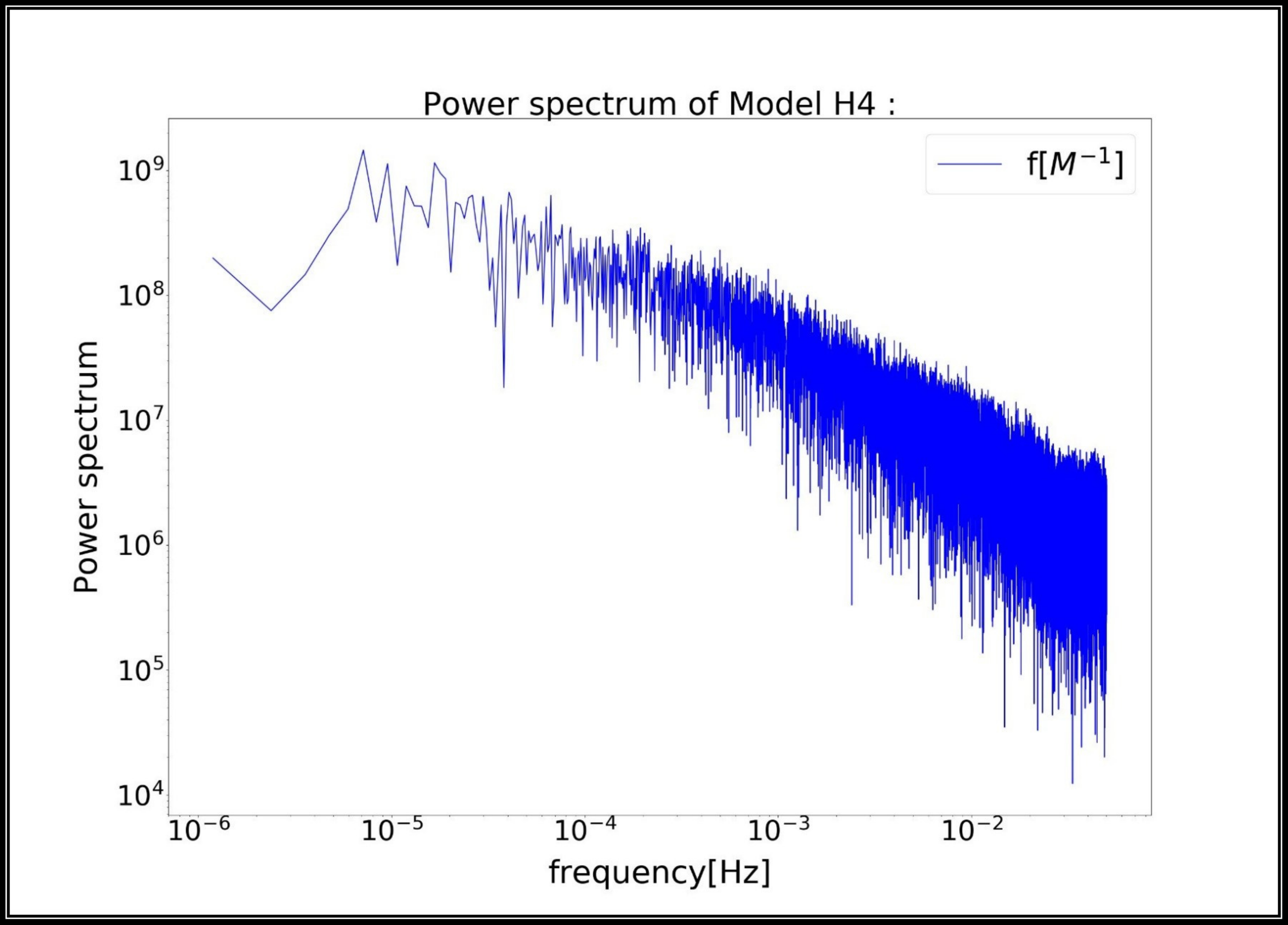}
     \caption{Model H4$[\gamma = 1.4, \lambda =3.6  \rm [M] , \epsilon = 0.0001 $, a = 0.10]. The figure shows the corresponding inner mass accretion rate of this model. See the values of frequency and other details in the text. }
     \label{fig:13}
 \end{figure}

    \begin{figure}
     \includegraphics[width = 0.5\textwidth]{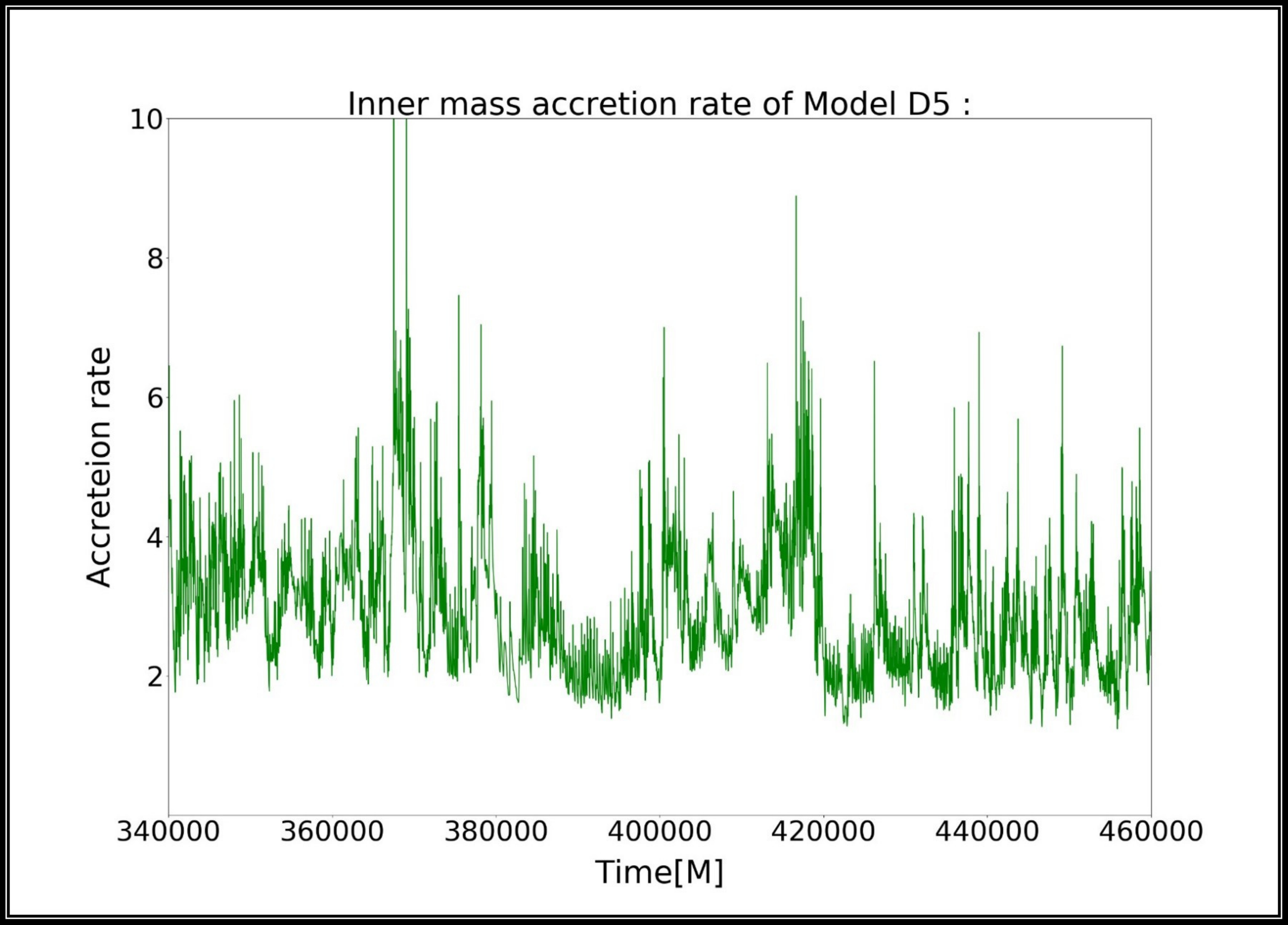}\\
     \caption{Mass accretion rate for model D5  $[\gamma$ = 1.4, $\lambda $=3.86 [\rm M], $\epsilon$ = 0.0001] . The x scale has been zoomed in to show the flaring in the accretion rate for model D5.}
     \label{fig:14}
 \end{figure}

   \begin{figure}
    \textbf{(a)}\\
     \includegraphics[width = 0.5\textwidth]{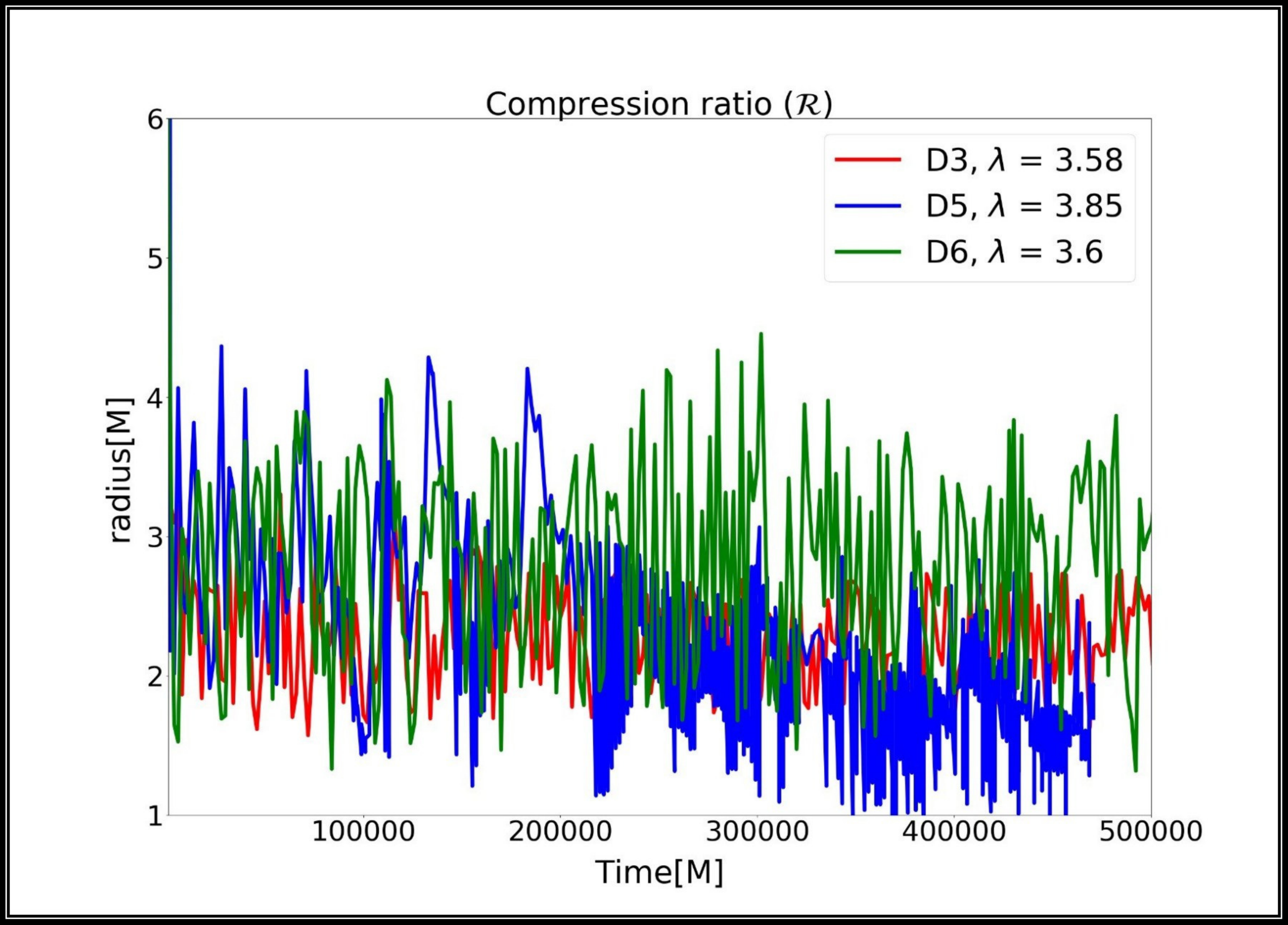}\\
     \textbf{(b)}\\
     \includegraphics[width = 0.5\textwidth]{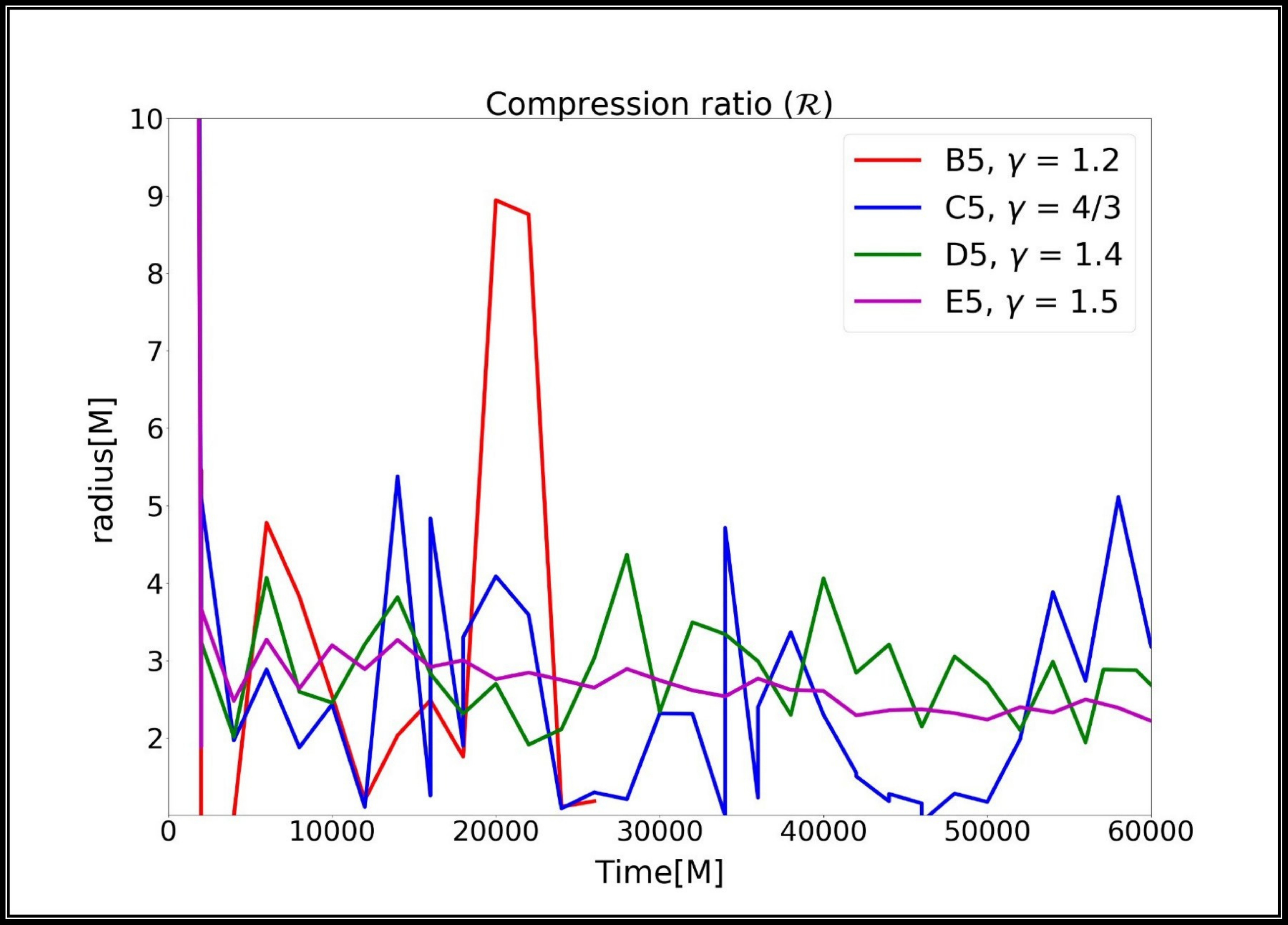}\\
     \caption{Panel (a) shows compression ratio varying with angular momentum $\lambda$ and energy $\epsilon$ over time for model D3 $[\gamma = 1.4, \lambda =3.58  \rm [M], \epsilon = 0.0005 ]$, D5 $[\gamma = 1.4, \lambda =3.86  \rm [M] , \epsilon = 0.0001 ]$ and D6 $[\gamma = 1.4, \lambda =3.6  \rm [M] , \epsilon = 0.0001 ]$. Panel (b) shows compression ratio varying with adiabatic index $\gamma$ over time for model C5 $[\gamma = 4/3, \lambda =3.86  \rm [M], \epsilon = 0.0001 ]$, D5 $[\gamma = 1.4, \lambda =3.86  \rm [M], \epsilon = 0.0001 ]$ and E5 $[\gamma = 1.5, \lambda =3.86  \rm [M], \epsilon = 0.0001 ]$.}
     \label{fig:15}
 \end{figure}
  Contrary to the situation for model A1 with $\gamma = 1.02 $, model F1 with $\gamma = 5/3$ is confined with subsonic flow. For gas with $\gamma = 5/3$, evolution can be thought of as completely dominated by gas pressure and rotation causes no effect to it for our values of $\epsilon$ and $\lambda$.  This result is in agreement with the work of \cite{proga2003accretion} where they show that for $\gamma$ = 5/3, the outer sonic point can only be found for a very slow rotation of gas at very large radii for zero total energy.

 The time variation of mass accretion differs significantly as $\gamma $ varies from $\gamma = 1.02$ to $\gamma = 5/3$. Accretion rate  has been observed either relatively constant or exhibiting small-amplitude quasi-periodic oscillations depending upon the adiabatic index.
   
For higher $\gamma$, the gas pressure is larger and the information about the perturbation propagates faster. The gas pressure works against effects of shock and tries to restore the uniformity of the flow. It can be seen that the non uniformity of the flow is initially induced by the numerical effects, but the propagation and amplification of the non uniformity strongly depends on the physical conditions in the accretion flow. In the cases where the gas pressure is reduced because of low $\gamma$, the non uniformity of the flow is stronger. It appears that for smaller $\gamma$ shock position is closer to the black hole since lower $\gamma $ means cooler disk with larger Mach number of the flow.

  For all presented models the outer boundary, which is set to 50000 [M] ($\sim 750000$ km for a 10 Solar mass black hole),
  is significantly farther away than the position of the outer sonic point.
  From models where shock expands till the outer sonic point, the most distant outer sonic point is located in the case of model E6, where it lies at 2789 [M] ($\sim 42000$ km). The inflow of matter from the outer boundary down to the outer sonic point is subsonic and the velocities of the flow far away from the center are very low. If the shock reaches the position of the outer sonic point, the two merge, leaving the flow subsonic all the way down to the inner sonic point (which is located very close to the black hole at a few [M]).
  In the subsonic flow, the shock no longer exists. Therefore we do not have any direct interaction between the shock front and the outer boundary. On the other hand, the (subsonic) inflow of matter is important because we want to study the long-term behaviour of the shock front in the flow, which is determined by the outer boundary condition.


 \subsubsection{Spinning black hole}
We computed a few models for spinning black hole, with $\gamma$ = 1.4. Only two values for spin has been chosen, one to be quite small (a= 0.10) and other to be quite large (a = 0.89). The models with non zero spin are named as H1[$\lambda$ = 3.86 [M], $\epsilon$ = 0.0001, a = 0.10],
H2[$\lambda$ = 3.86 [M], $\epsilon$ = 0.0025, a = 0.10], H3[$\lambda$ = 3.86 [M], $\epsilon$ = 0.0001, a = 0.89], H4[$\lambda$ = 3.6 [M], $\epsilon$ = 0.0001, a = 0.10], H5[$\lambda$ = 3.6 [M], $\epsilon$ = 0.0025, a = 0.10] and H6[$\lambda$ = 3.6 [M], $\epsilon$ = 0.0001, a = 0.89] in the Table [\ref{Table 1}]. 

It is interesting to see that shock oscillates quite nicely for model H4 (with $\epsilon$ = 0.0001) but not for H5 (with $\epsilon$ = 0.025) where $\epsilon$ increases and shock gets accreted. (See Figure [\ref{fig:13}] showing the oscillation of the shock front for model H40). Previous article \citep{sukova2017shocks} presented quite in detail the accreting or oscillating shock depending on the spin of the black hole for $\gamma $ = 4/3.
Here models H3 and H6 show expansion of the shock through the outer sonic point for higher value of $\lambda$ = 3.86 [M]. Though in many other astrophysical scenarios it can be seen that for increasing value of spin of black hole, the shock exists for even smaller values of $\lambda$
(see for example \citep{janiuk2018accretion}).

 
\subsubsection{ Frequency and power spectrum}
 \label{sec:frequency}
 \begin{table}
    \centering
    \begin{tabular}{c|c|c|c}
         \large{Models}& \large{Mean of ($\dot{  \rm M}$)}& \large{Frequency (M$^{-1}$)}& \large{Amplitude}\\ [2ex] 
  \hline  
B5& 4.2*10$^{5}$& 10$^{-3}$& 2.67 \\    
D3& 3.7*10$^{5}$& 3.7*10$^{-4}$; 2.6*10$^{-4}$& 0.32\\  
D5& 2.9*10$^{5}$& 1.2*10$^{-4}$& 3.79\\
D6& 3.6*10$^{5}$& 3*10$^{-5}$& 1.06\\    
H4& 3.5*10$^{5}$& 10$^{-5}$& 5.06\\
    \end{tabular}
    \caption{Column 1 of the table mentions the name of the model, column 2 is for the mean value of accretion rate. Here accretion rate is in code units. The oscillation frequency and corresponding amplitude of oscillation are in column 3 and 4 respectively. The models where oscillation is significant, only those are mentioned in the table and the frequencies mentioned are an estimated inference from PDS and shock position variation (see text for explanation).}
    \label{Table 2}
\end{table}
The results from frequency analysis for the selected models have been tabulated in Table [\ref{Table 2}].  Our simulation covers only quite short segment of the hypothetical light curve (about 50s for typical microquasar), so we did not proceed with detailed and elaborate time series analysis. Instead we only perform fast Fourier transform on the data series of $\dot{\rm M}(t)$ and look for indication of peaks in the spectrum  corresponding to the vertical oscillations of the shock bubble. Amplitude being calculated as [($\dot{\rm M}_{max}$ - $\dot {\rm M}_{min}$) / $\dot{\rm M}_{mean} $] gives an insight whether its a small scale oscillation or a large scale oscillation.

Model B5 shows oscillation with quite high amplitude (2.67) during short time interval. At $t\approx 26400$\,M the shock is accreted and the accretion flow follow the supersonic branch of solution from the outer sonic point down to the black hole.
Two faint peaks can be seen  in the spectrum with frequency $\sim$ 1.0*10$^{-3}$[M$^{-1}$] and 2.8*10$^{-3}$[M$^{-1}$].

Model D3 can be seen in Figure [\ref{fig:6}(b)] showing quite nice long scale oscillation. The amplitude is quite small. Two peaks in the spectrum has been observed at frequency 3.7*10$^{-4}$ [M$^{-1}$] and 2.6*10$^{-4}$ [M$^{-1}$]. These peaks are quite close to the observed twin peaks quasi periodic oscillation of ratio 3:2.

Figure [\ref{fig:14}] shows the zoomed in mass accretion rate for model D5 which shows the irregular flaring state in the light curve. To see a QPO in this model, a longer data set would be needed.

The inner mass accretion and the oscillation of the shock position in time has been shown in the Figure [\ref{fig:12}] for model D6. Though this model shows quasiperiodic oscillation, again the data set is too short  to obtain the signal of the QPO in the power spectrum (it covers about 50s for a 10 M$_{\odot}$  black hole, which is typical mass of a microquasar). However, in real observed sources, the QPOs are often quite weak and requires to combine more observations with duration of hundreds or thousands of seconds of the same source in the same spectral state together to find the corresponding peak in the spectrum. Hence, it is beyond the computational scope of the presented paper to perform long enough simulations to obtain the accurate frequency of the QPO. 
 
Model H4 also shows significant oscillation of shock position. Several very pronounced peaks with duration about 10$^{4}$ [M] (which correspond to $\sim$ 0.5s
 can be seen in the accretion rate in Figure \ref{fig:13}, which are accompanied by large oscillation of the shock bubble.  The amplitude of oscillation is quite high as 5.06. The interval between the peaks varies on the order of hundred thousand M, from which we can roughly estimate the frequency as 10$^{-5}$ [M$^{-1}]$, which corresponds to 0.2 Hz for 10 solar mass black hole.

   We note that it is hard to determine the QPO frequency from the direct
   inspection of the Power density spectra (PDS) derived in our model,
   as shown in Figures [\ref{fig:5}], [\ref{fig:6}], [\ref{fig:12}], [\ref{fig:13}] and [\ref{fig:14}].
   Nevertheless, we argue that the oscillation of shock front in our
   computation has a quasi-periodic nature, as revealed by the time dependence of its equatorial  size.
The shock position for model H4 is shown in the  Figure [\ref{fig:13}].
During the span of the simulation, only about 10 cycles happen, which is
too little number for the peak to appear pronounced on the PDS.

Those oscillations can be attributed to two
processes. First, we can see vertical oscillations of the shock bubble.
This causes faster oscillations of the accretion rate with quite small
amplitude and small motion of the shock front in the equatorial plane. It can be ascribed to the mixing of low and high angular momentum
gas along the boundary of the funnel, where the steepest slope of the
angular momentum distribution occurs (see the 2D map of angular momentum, e.g. in Figure [\ref{fig:4}]. This effect was already seen in previous
studies of similar cases, e.g. by \citet{MoscibrodzkaProga2008} for
pseudo-Newtonian computations of similar setup.
Second, this mixing occasionally happens not only at the boundary of
the funnel, but the low angular momentum gas flows through the outer
part of the shock bubble and even crosses the equator (see panel (a) on Figure [\ref{fig:11}] ).
When this
happens, the physical conditions around the shock change (suddenly
there is a lot of gas with a much lower angular momentum), the shock
shrinks towards the black hole (Figure [\ref{fig:11}(b)] ) and this is accompanied by the increase
of the accretion rate, when part of the shock bubble is accreted.
When
the low angular momentum gas is accreted from the equatorial region, the
conditions restore and the shock bubble is rebuilt. The shock front
moves farther from the black hole and the accretion rate decreases (Figure [\ref{fig:11}(c)] ). This
process happens with a longer period and in the time span  of our
simulation we can see only a few of those events for each model.
However, when
they develop, they provide larger peaks in the accretion rate. It can be seen in the Figure [\ref{fig:13}(a)] that the mass accretion rate shows larger peaks around t $\sim$ 500000[M] as the shrinking shock rebuilds.

Those episodes of sudden shrinking of the shock bubble are important for
the shape of the total light curve, as seen from the accretion rate
variation, but they probably also influence the spectrum of the outgoing
radiation. In particular, it is reasonable to assume, that substantial part of the
radiation comes from the shock region, and hence the conditions at the shock front influence the energy dissipation.
Because the shock front is significantly changing its
position (in case of model H4 it oscillates between r$_{s}$ = 60[M] and 220 [M]), those
parameters (e.g. density and temperature) are also changing, which can lead to shifts in the spectral energy distribution. See Figure [\ref{fig:11}].


 \section{Discussion}
 \label{sec:discussion}
 Transonic black hole accretion has been studied in detail over past 30 years \citep{paczynski1981model, kumar1990theory, das2012hysteresis}.  Here we discuss the results from our simulation in context of former analytical and numerical works. Also we show some of the implications in astrophysical scenario.

\subsection{Correspondence to former models}

 The location and behaviour of shock front with respect to $\lambda$ and $\gamma$ in our results from hydrodynamical simulation agrees with the range presented in the  article \cite{sukova2015shocks} (see Figure 6 from that article where the variation of shock position with respect to $\gamma$ and $\lambda$ was shown using ZEUS code \citep{stone1992zeus} with Paczynski-Wiita potential). The shock position can be seen little more pushed further away from black hole in our GR simulations, which can be seen e.g. for model D6 with $\gamma=1.4, \lambda=3.6$  [\rm M] and $\epsilon=0.0001$. For these parameters the earlier result with Paczynski-Wiita potential predicted the shock position at about $r_s=50$ [\rm M], while our GR simulation show the shock oscillating between 44  [\rm M] to 98 [\rm M] with mean shock position at 71  [\rm M]. Moreover, for $\gamma=1.4, \lambda=3.86$  [\rm M] and $\epsilon=0.0001$ the plot in Fig. 6 of \cite{sukova2015shocks} predicts non-existence of the shock front, while we see in model D5 the shock doing small oscillations and then expanding in the range 10 to 3790 M with mean position around 1900 M.
  Also the minimal and maximal shock position r$_{s}^{min}$ and r$_{s}^{max}$ evolves in accordance with \cite{sukova2015shocks} and gets farther as the value of adiabatic index increases. We get the difference in position of  r$_{s}^{min}$ and r$_{s}^{max}$  for model B5 as 77.6 [\rm M], for model C5 as 291 [\rm M] and for model D5 as 3780.4 [\rm M].

  We have also compared the frequency we get here from the simulation to that calculated from analytic solution and they are in quite good agreement.
   Notice that these analytic solutions are following from the
    stationary model, and hence 'frequency' is computed simply from the free-fall timescale at the shock radius, which in fact does not oscillate. 
    
  We choose models D6 and H4 for comparing numerical and analytical frequency as these models show significant peak during the evolution of the flow. The frequencies estimated for model D6 and H4 are  2.97*10$^{-5}$ [M$^{-1}$] and  10$^{-5}$ [M$^{-1}$] respectively (see Table \ref{Table 2}). These values corresponds to  frequency of 0.6 Hz and 0.2 Hz  for a 10 M$_{\odot}$ black hole. To compare with the analytical solution, we used the formulae for the QPO frequency presented in  \citep{iyer2015determination, chakrabarti2008evolution}.
 \begin{equation}
     \nu_{QPO} = \frac{c/r_{g}}{2 \pi  \mathcal{R}  r_{s} \sqrt{r_{s}-1}} 
 \end{equation}
 where r$_{g}$ is $2G M_{BH}/ c^{2}$.
 Using this relation we get the frequency value as 0.57Hz for compression ratio 3.01 and shock position r$_{s}^{max}$ = 97.5 [M] obtained from the simulation of model D6 for a 10 M$_{\odot}$ black hole. Similarly for model H4, we obtain a frequency value of 0.22Hz for compression ratio 2.34 and shock position r$_{s}^{max}$ = 219.6 [M].
 In our case the shock responds to the physical conditions in the flow and moves, and simultaneously reflects also the magnitude of compression ratio $\mathcal{R}$ changes. The changes of values of shock position and compression ratio are quite substantial (see Figure [\ref{fig:15}] for the evolution of the compression ratio during the simulation), so using the analytic relation given by the stationary model, which does not take this into account, can provide only rough  estimates of the frequency.

 A more recent semi-analytic work shows the effects of variable adiabatic index on shock formation and oscillation by using a relativistic EoS \citep{dihingia2019low}. We note that in our code we use the adiabatic EoS with a constant $\gamma$, due to the difficulties of the conservative MHD scheme embedded in HARM.
 Nevertheless, our results stay in agreement with this work for their models with Ideal EoS (IEoS) and spinless black hole with a = 0. As it is shown in \cite{dihingia2019low}, for different values of $\epsilon$, $\lambda$ and spin parameter $a$, the range of compression ratio covers the values of ($\mathcal{R}$) as $\mathcal{R}_{max} = 3.79$ down to $\mathcal{R}_{min} = 1.19$.  From our simulation, we get a $\mathcal{R}_{max} = 3.88$ and $\mathcal{R}_{min} = 1.3$ for model D6 with $\gamma$ = 1.4 during the whole evolution of the flow. The compression ratio corresponding to maximal and minimal shock position for this model D6 is 3.01 and 2.45 correspondingly.
Moreover, we can confirm their statement, that with constant $\gamma$ and $\epsilon$, the compression ratio decreases with increasing angular momentum of the flow. This trend is seen in all pairs of our corresponding models: D6 ($\lambda=3.6$ \rm [M], $\mathcal{R}=3.2$) and D5 ($\lambda=3.86$ \rm [M], $\mathcal{R}=1.7$); D3 ($\lambda=3.58$ \rm [M], $\mathcal{R}=2.9$) and D4 ($\lambda=3.72$ \rm [M], $\mathcal{R}=1.8$); E6 ($\lambda=3.6$ \rm [M], $\mathcal{R}=2.9$) and E5 ($\lambda=3.86$ \rm [M], $\mathcal{R}=2.7$).

In Figure \ref{fig:15}  we present the time dependence of the compression ratio $\mathcal{R}$ at the shock front for several models. The compression ratio varies as the shock moves towards and outwards from the black hole, in particular the value of compression ratio anti-correlates with the shock position, as expected. The range of the variation is quite high, e.g. in model C5 the value ranges from 1 up to 5. Also the trend is, that models with lower energy exhibit larger variation of $\mathcal{R}$.

\subsection{Astrophysical significance}

 Our results may be important for the observed
 low frequency QPO's (LFQPO) in microquasars.
 Microquasars have been observed doing oscillations in range of few hundreds mHz up to few tens of Hz \citep{markwardt1999variable, cui1999phase, remillard1999rxte, nandi2012accretion}.
These LFQPO's have been observed  in the frequency range of 0.05Hz - 10Hz in many x-ray binaries and microquasars such as GRO J1655-40 \citep{vignarca2003tracing, remillard1999rxte}, XTE J1118-480 \citep{revnivtsev2000discovery, wood2000usa}, XTE J1748-288 \citep{revnivtsev2000rxte, sobczak2000correlations}, IGR J17091-3624 \citep{iyer2015determination, altamirano2012low}.
We found that several models from our simulations show subtle and definitive shock front oscillations over time such. This is found in models D3, D6 and H4
where the value of oscillation frequency obtained  with our simulations
is representative for the time variability found in the above microquasars.

The Galactic black hole candidate IGR J17091-3624, as well as
GRS 1915+105, are also displaying wide range of temporal and spectral variations.
The luminosity is much less (about 50 times) in IGR J17091-3624 than the one observed for GRS 1915+105 \citep{altamirano2011faint},
however there is a wide discussion
about the mass of this black hole candidate \citep{rodriguez2011first, rao2012igr, iyer2015determination}. We choose the mass range 8.7M$_{\odot}$ - 15.6 M$_{\odot}$ for this black hole \citep{iyer2015determination} for which
we get the estimated oscillation
frequency range between 0.6Hz and 0.3Hz (from model D6),
between 0.24Hz and 0.13Hz (from model H4), and between
8.6Hz and 4.8Hz (from model D3).
Our frequency range fits broadly into the observed frequency of this source (0.005Hz - 5Hz). Our simulation results are also in good agreement with the frequency range (0.1Hz - 15Hz) observed in black hole system GRO J1655-40.
This shows that our model could be a good explanation for LFQPO's from these sources, under the assumption that the oscillations come from the inner parts of
the adiabatic flow described with $\gamma=1.4$ that accretes onto a non-spinning, or only a moderately spinning black hole. Further investigation is needed
to study in detail the range of spins of the Kerr black holes that
is able to produce an oscillatory shock behaviour in the low
angular momentum accretion flows.

  Finally, our investigations of the shock front oscillation may be relevant for some of the observed Active Galactic Nuclei. Here the observations are not as definitive as in the case of Galactic X-ray binaries, but the combined constraints from the energy spectrum and variability show that the soft excess is likely arising from the low-temperature Comptonization of the disc. This remains more or less constant on short time-scales, diluting the QPO and rapid variability seen in the power-law tail of the Seyfert galaxy RE J1034+396 \citep{Middleton2009}.
  Also, after careful modeling of the noise continuum, the $\sim 3.8$ hr QPO
  was found in the ultrasoft AGN candidate 2XMM J123103.2+110648 \citep{Lin2013}. The tentative detection might suggest that the shock front in this AGN oscillates in
  several modes (equatorial, polar, azimuthal), as suggested by our results.

 \section{Acknowledgement}
 We thank Konstantinos Sapountzis and Ireneusz Janiuk for helpful discussions and computational support.
We also thank Giovanni Miniutti for helpful comments.
 This work was partially supported by the grant No. DEC-2016/23/B/ST9/03114 from the Polish National Science Center. We also acknowledge support from the Interdisciplinary Center for Mathematical Modeling of the Warsaw University, through the computational grant Gb70-4, and the PL-Grid computational resources through the grant grb2. PS is supported from Grant No. GACR-17-06962Y from Czech Science Foundation.


\label{lastpage}

\end{document}